\documentclass[preprint,12pt]{elsarticle}

\usepackage{tikz}
\DeclareRobustCommand{\redcircles}{%
    \begin{tikzpicture}
        \filldraw[red] (0,0) circle (0.6ex); % Adjust size by changing 1ex
    \end{tikzpicture}%
}
\DeclareRobustCommand{\greentriangle}{%
    \begin{tikzpicture}[baseline=-0.8ex]
        \draw[green] (0,0) -- (0.1cm,0.18cm) -- (0.2cm,0) -- cycle;
    \end{tikzpicture}%
}
\DeclareRobustCommand{\bluediamond}{%
    \begin{tikzpicture}[baseline=-0.5ex]
        \draw[blue] (0cm,0.1cm) -- (0.1cm,0) -- (0.2cm,0.1cm) -- (0.1cm,0.2cm) -- cycle;
    \end{tikzpicture}%
}
\DeclareRobustCommand{\orangesquare}{%
    \begin{tikzpicture}[baseline=-0.5ex]
        \draw[orange] (0,0) rectangle (0.2cm,0.2cm);
    \end{tikzpicture}%
}

\DeclareRobustCommand{\trianglewithline}{%
    \begin{tikzpicture}[baseline=-0.5ex]
        \draw[red] (0,0) -- (0.1cm,0.2cm) -- (0.2cm,0) -- cycle;
        \draw[red] (-0.1cm,0.1cm) -- (0.3cm,0.1cm);
    \end{tikzpicture}%
}

\usepackage{subfigure}
\usepackage{caption}
\usepackage{amssymb}
\usepackage{amsmath}
\usepackage[ruled,vlined]{algorithm2e}
\usepackage{booktabs}
\usepackage[most]{tcolorbox}  % Load tcolorbox with most libraries
\usepackage{soul}             % For highlighting text

\newtcolorbox{dialogbox}{
  arc=4mm,
  colback=blue!3,
  colframe=black,
  rounded corners,
  boxrule=0.5pt,
  fonttitle=\bfseries,
  coltitle=black,
}

\usepackage{adjustbox}

\definecolor{lightblue}{HTML}{ADD8E6}
\definecolor{vividlightblue}{HTML}{A0D8EF}
\definecolor{brighterlightblue}{HTML}{B3E2F2}
\definecolor{lightorange}{HTML}{FFA07A}
\definecolor{lightred}{HTML}{FFB499}
\definecolor{lightgreen}{HTML}{B7F4B7}
\definecolor{lightgrey}{HTML}{D3D3D3}

\newcommand{\komegasst}{$k$-$\omega$ SST}

\def\ourmethod{AutoTurb}

\begin{document}

\begin{frontmatter}


\title{AutoTurb: Using Large Language Models for Automatic Algebraic Model Discovery of Turbulence Closure}

\author[a]{Yu Zhang}
\author[b]{Kefeng Zheng}
\author[b]{Fei Liu\corref{cor1}}
\cortext[cor1]{Corresponding author}
\ead{fliu36-c@my.cityu.edu.hk}
\author[b]{Qingfu Zhang}
\author[a]{Zhenkun Wang} %% Author name
\affiliation[a]{organization={School of System Design and Intelligent Manufacturing, Southern University of Science and Technology},
            city={Shenzhen},
            postcode={518055},
            country={China}}
\affiliation[b]{organization={Department of Computer Science, City University of Hong Kong},
            city={Hong Kong},
            country={China}}
%% Abstract
\begin{abstract}
Symbolic regression (SR) methods have been extensively investigated to explore explicit algebraic Reynolds stress models (EARSM) for turbulence closure of Reynolds-averaged Navier-Stokes (RANS) equations. The deduced EARSM can be readily implemented in existing computational fluid dynamic (CFD) codes and promotes the identification of physically interpretable turbulence models. The existing SR methods, such as genetic programming, sparse regression, or artificial neural networks, require user-defined functional operators, a library of candidates, or complex optimization algorithms. 
Recently, large language models (LLMs), trained on large amounts of publicly available source code, have drawn great attention for their abilities to generate computer programs with more general free-text inputs and problem descriptions, and provide wider possibilities with novel insights.
In this work, a novel framework using LLMs to automatically discover algebraic expressions for correcting the RSM is proposed. The direct observation of Reynolds stress and the indirect output of the CFD simulation are both involved in the training process to guarantee data consistency and avoid numerical stiffness. Constraints of functional complexity and convergence are supplementally imposed in the objective function on account of the tremendous flexibility of LLMs. The evolutionary search is employed for global optimization. 
The proposed method is performed for separated flow over periodic hills at $Re = 10,595$. The generalizability of the discovered model is verified on a set of 2D turbulent separated flow configurations with different Reynolds numbers and geometries. 
It is demonstrated that the corrective RANS can improve the prediction for both the Reynolds stress and mean velocity fields. Compared with algebraic models discovered by other works, the discovered model performs better in accuracy and generalization capability. The proposed approach provides a promising paradigm for using LLMs to improve turbulence modeling for a given class of flows.
\end{abstract}

%\item A novel framework using LLMs to automatically discover algebraic expressions for correcting the turbulence closure is proposed. 
%\item A general natural language-based prompt is put forward for learning the turbulence data.
%\item The direct observation of Reynolds stress and the indirect output of the CFD model are both incorporated to guarantee data consistency. 
%\item Constraints of functional complexity and simulation convergence are imposed to avoid divergence or ill-conditioning and ensure a stable optimization process.
%\item Compared with algebraic models discovered by other works, the discovered model performs better in accuracy and generalization capabilities.

%% Keywords

\end{frontmatter}



\section{Introduction}
\label{sec1}
%% Labels are used to cross-reference an item using \ref command.

Computational fluid dynamic (CFD) methods including Reynolds-averaged Navier-Stokes (RANS)~\cite{Pope2000RANS}, large-eddy simulations (LES)~\cite{Ferziger1985}, and direct numerical simulations (DNS)~\cite{Moin1998DNS} have become promising ways to model and study fluid mechanism in the past decades. Among them, the RANS method has gained widespread applications in industrial areas such as aerospace engineering due to its lower computational cost and superior robustness. However, for flow with separations, strong pressure gradient, and curvature, the RANS method frequently fails to deliver satisfactory simulation outcomes due to the linear eddy-viscosity assumption and semi-empirical turbulence modeling~\cite{Wilcox1998Turbulence}. 
Therefore, improving turbulence modeling in RANS to achieve accurate predictions of complex flow behaviors is crucial for engineering applications.
   
In recent years, thanks to the rapid development of high-performance computing architectures as well as increasing accessibility of high-fidelity flow data, data-driven approaches like machine learning (ML) techniques have emerged as promising techniques for physical modeling~\cite{Duraisamy2018Turbulence}. 
These technologies provide a new alternative in the analysis and understanding of turbulent flows and become a new paradigm for correcting the governing equations~\cite{Yan2022Data,Wang2016Physics,Singh2016Machine}, uncertainty quantification~\cite{Cheung2011Bayesian,Platteeuw2008Uncertainty,Xiao2017A,Margheri2014Epistemic} and constructing new constitutive models for Reynolds stress tensors~\cite{Gamahara2016Searching, Zhu2019Machine,Yin2020Feature,Jiang2021An}.
These researches significantly improve the performance of RANS equations in complex situations. However, the black-box nature of most ML methods hampers the understanding of obtained data-driven models, making it difficult to provide physical interpretations and infer new flow physics. The limited interpretability and generalizability prevent these black-box models from being accepted by the engineering community and increase the difficulty of disseminating the learned models to end users since there are no explicit mathematical equations.
   
In order to solve the aforementioned problems associated with black-box ML models, symbolic regression (SR) based data-driven turbulence modeling methods have been proposed and gained a renaissance in the ML community for finding equations from sparse data~\cite{Chen2021Physics}. The task of symbolic regression is to find an explicit algebraic expression that best predicts the target given input variables. Once the expression is derived, it can operate independently of the training environment, making it easier to integrate into existing CFD solvers with a negligible increase in computational cost. Besides, researchers can integrate relevant functional forms and physical quantities into the SR based on turbulence laws and physical insight, ensuring that the final corrected method possesses greater physical significance.

According to the search method used for SR in the field of turbulence modeling, it can be classified into four categories: 1) The genetic programming (GEP) method, which is first introduced by Weatheritt J. et al.~\cite{Weatheritt2016A}, uses an evolutionary algorithm (EA) to find the global optimum tensor regression for the anisotropy of Reynolds stress. The candidate solution is encoded as a linear string of predefined operators and variables. A CFD-driven approach is developed using any flow feature from the CFD results to train models~\citep{ZHAO2020109413}. Then, a multi-objective evolutionary algorithm is proposed to combine multiple training objectives~\citep{WASCHKOWSKI2022110922}. 2) The SpaRTA method, proposed by Schmelzer et al.~\cite{Schmelzer2019}, implements a sparse regression using regularization terms for correcting Reynolds stress anisotropy and production of transported turbulent quantities resolved by a frozen-RANS method. A library of candidates and operators should be predefined. 3) a CFD-driven symbolic identification method (CFD-driven SpaRTA) is further developed~\citep{BENHASSANSAIDI2022}, which uses the surrogate-based method to find the optimum expression. 4) The deep SR (DSR) method proposed by Tang et al.~\cite{Tang2023SR}, combining deep learning with SR, employs a risk-seeking policy gradient algorithm to train neural networks to generate the best expression for the RSM.
Besides the searching method, the main characteristics of these methods are summarized in Table~\ref{tab: methods}. It is shown that predefined operators or functional candidates are required by the existing methods. Functional complexity is the most important factor in SR methods to obtain pragmatic and interpretable models. In GEP, the length and number of genes are defined to restrict the complexity. In SpaRTA and CFD-driven SpaRTA methods, Lasso- and Ridge-regressions are employed to promote sparsity and relatively small coefficients. The minimal and maximal lengths of expressions are pre-specified in the DSR method and handled in the prior sampling stage. 
The current methods require the researchers to be highly experienced when designing new models and demand tremendous human effort.

\begin{scriptsize}
\begin{table} [ht]
    \centering
    \small
    \caption{Key features of existing SR methods and our approach.}
    \label{tab: methods}    
    {
        \begin{tabular}{p{60pt}p{110pt}p{80pt}p{80pt}}
        \toprule[1pt]
        Methods & Expression & Search method & Constraints \\
        \hline 
        \specialrule{0pt}{1pt}{1pt}
       GEP~\cite{Weatheritt2016A}    & chromosome encoded as linear string & EA & -  \\
        \specialrule{0pt}{1pt}{1pt}
       SpaRTA~\cite{Schmelzer2019} & library of candidates with coefficients& linear regression & complexity \& robustness\\
        \specialrule{0pt}{1pt}{1pt}
       \small {CFD-driven SpaRTA}~\citep{BENHASSANSAIDI2022} & library of candidates with coefficients & surrogate-based method & complexity \& robustness \\
        \specialrule{0pt}{1pt}{1pt}
       DSR~\cite{Tang2023SR} & sampled token of operator and variables & gradient-based method & - \\
       
        \specialrule{0pt}{1pt}{1pt}
       AutoTurb (Ours) & arbitrary form and length & EA + LLMs & complexity \& convergence  \\
       \bottomrule[1pt]
       \end{tabular}
    }
\end{table}      
\end{scriptsize}

In the past years, large language models (LLMs)\cite{naveed2023comprehensive}, which are trained on large amounts of publicly available source codes, have become increasingly powerful due to the intensive training data and large model sizes. Pre-trained LLMs show impressive language processing and code generation capability. It has drawn great attention in algorithm design~\cite{liu2023algorithm,yao2024evolve}, mathematical discoveries~\cite{romera2024mathematical} and optimization~\cite{guo2023towards,zhang2023using,mirchandani2023large}. 
Recent works have shown that in comparison to standalone LLMs, incorporating LLMs in a search framework could significantly boost the performance in some hard reasoning and designing tasks~\cite{liu2024evolution,yao2024evolve}. The application of LLMs in scientific discoveries offers a versatile and automated approach, significantly reducing the need for human intervention and specialized expertise during the design process.

In this paper, we present the first attempt to adopt LLMs for turbulence model discovery of the RANS method, as shown in Table~\ref{tab: methods}. LLMs are employed to formulate the optimization problem using natural language, and autonomously generate algebraic expressions for correction models.
Research has demonstrated that when correction models are trained offline using high-fidelity Reynolds stress fields, the modified RANS equations could be ill-conditioned~\cite{Duraisamy2021}. Even when Reynolds stresses with errors under 0.5$\%$ from direct numerical simulation (DNS) databases are substituted into RANS equations, the resulting velocity fields can exhibit significant errors (up to 35$\%$)~\citep{Wu2019}.
To maintain data consistency and avoid numerical instability, a CFD-driven approach is employed, with constraints placed on the complexity of the generated expressions. Moreover, the vast design freedom provided by LLMs necessitates strict constraints on numerical convergence in CFD to ensure a stable optimization process.
Applications in different numerical experiments including Periodic Hills with different geometry and Reynolds numbers, converging-diverging channels, and curved backward-facing step were used to validate the generability of the proposed method. 

%\red{huge DoF, conditioning, numerical stiffness}

%FISR method\cite{Wu2023Enhancing}, combining SR with the field inversion framework, employs a gradient-based method to derive the symbolic mapping between the local flow features and the corrective field. Stocker et al.\cite{YvonneStcker2024dns} enhanced the RANS simulation of two-phase particle flow using the sparse SR approach. He et al.\cite{He2024Afield} utilized SR to establish an analytic expression mapping local flow field variables to the SA model correction factor, and improved the accuracy for airfoil stall prediction. Song et al.\cite{Song2024Aninnovaaive} employed the symbolic regression (SR) to scrutinize the physical correlation between Pk and local turbulence parameters, improving the consistency with high-accuracy data and experimental results in early separation problems. Li et al.\cite{Li2024Evolutionary} combined one evolutionary algorithm, gene expression programming (GEP), with an artificial neural network (ANN) for symbolic regression. 

The structure of this paper is as follows: In Section 2, we introduce the governing equations of the RANS model and the mathematical formulation of \komegasst{} turbulence model. We also introduce basic assumptions about the nonlinear Reynolds stress model used in this paper. In Section 3, the proposed framework based on LLMs, named AutoTurb, is provided. Numerical experiments of training and cross-validation are present in Section 4 and Section 5. In Section 6, we conclude this paper and outline future research directions. 
   
\section{Preliminary}

This section briefly recalls preliminary methodologies including governing equations of incompressible RANS, the baseline linear viscosity model combined with \komegasst{} turbulence model, and the basis of nonlinear algebraic correction for RSM.
  
\subsection{RANS with turbulence modeling}

The steady incompressible RANS equations are
\begin{equation} \label{RANS}
\begin{aligned}
    \frac{\partial U_i}{\partial x_i} &= 0, \\ 
    U_j \frac{\partial U_i}{\partial x_j} &= \frac{\partial}{\partial x_j} \left[ -\frac{P}{\rho} + \nu \frac{\partial U_i}{\partial x_j} + \tau_{ij} \right],
\end{aligned}
\end{equation}
\noindent where $U_i$ with $i \in \lbrace 1,2,3 \rbrace$ and $P$ are components of the mean-flow velocity and the mean pressure. $\rho$ and $\nu$ are constant density and the kinematic viscosity. The instantaneous fluctuation of turbulence is represented by a Reynolds stress tensor $\tau_{ij}$ in the momentum equation. Known as the Boussinesq assumption, the anisotropic part of Reynolds stress $a^B_{ij}$ can be modeled by a linear constitutive relationship with the mean-flow straining field:  
\begin{align} \label{LEVM}
    \tau_{ij} = a^B_{ij} -\frac{2}{3} \rho k \delta_{i j} = 2 \nu_t S_{i j}-\frac{2}{3} \rho k \delta_{i j},
\end{align}
\noindent where $\nu_t$ is the turbulent kinematic viscosity or eddy viscosity, $k:= \frac 1 2 \tau_{ii}$ is the turbulence kinetic energy (TKE), and $S_{i j}$ is the trace-less mean strain rate tensor, which can be written as $S_{ij}=\frac{1}{2}\left(\frac{\partial U_i}{\partial x_j} + \frac{\partial U_j}{\partial x_i}\right)$ for an incompressible flow. 

The eddy viscosity $\nu_t$ can be computed from two transported variables, such as $k$ and the specific turbulence dissipation rate $\omega$. In this work, the popular \komegasst{} model is adopted as the baseline turbulence model, which defines transport equations for $k$ and $\omega$:
\begin{align}  \label{kw SST}
    \frac{\partial k}{\partial t} + U_j \frac{\partial k}{\partial x_j} &= \underbrace{\tau_{ij} \frac{\partial U_i}{\partial x_j}}_{P_k} - \beta^* k\omega + \frac{\partial}{\partial x_j} \left[(\nu + \sigma_k\nu_t) \frac{\partial k}{\partial x_j} \right], \\
    \frac{\partial \omega}{\partial t} + U_j \frac{\partial \omega}{\partial x_j} &= \frac{\gamma}{\nu_t}P_k - \beta\omega^2 + \frac{\partial}{\partial x_j} \left[(\nu + \sigma_\omega\nu_t) \frac{\partial \omega}{\partial x_j} \right] + 2(1-F_1)\sigma_{\omega2}\frac{1}{\omega}\frac{\partial k}{\partial x_i}\frac{\partial \omega}{\partial x_i} \ .
\end{align}
Then, the eddy viscosity $\nu_t$ is modeled as
\begin{align} 
    \nu_t := \frac{a_1k}{\max(a_1\omega, SF_2)} \ .
\end{align}

All remaining terms and coefficients are omitted for brevity, see \citep{Menter1994} for details.
The \komegasst{} model takes both the wall treatment and free-stream turbulence properties into account. It usually shows good performance in adverse pressure gradients and separating flow. However, because of the linear eddy viscosity assumption and TKE equilibrium assumption in the boundary layer, it is found that the \komegasst{} model produces unsatisfactory prediction of flow separation and the often underestimation of Reynolds stress.

\subsection{Nonlinear algebraic correction for Reynolds stress model}

The correction of the linear constitutive relationship for RSM is implemented based on an explicit nonlinear framework proposed by Pope \citep{Pope1975}. It is assumed that RSM depends not only on the strain rate tensor but also on the rotation rate tensor $\Omega_{ij}=\frac{1}{2}\left(\frac{\partial U_i}{\partial x_j} - \frac{\partial U_j}{\partial x_i}\right)$. Then, the most general form of the correction for non-dimensional RSM anisotropy is derived as a polynomial of ten basis tensors and five Galilean invariances:
\begin{align}\label{CHtheroem}
    \frac{a^\Delta_{ij}}{2k} = b^\Delta_{ij} (\tilde{S}_{ij}, \tilde{\Omega}_{ij}) = \sum_{n=1}^{10} T_{ij}^{(n)} \alpha_n(\lambda_1,\dots,\lambda_5),
\end{align} 
where $T_{ij}^{(n)}$ are basis tensors and $\alpha_n(\cdot)$ are arbitrary scalar functions of invariants $\lambda_m, m=1,\dots,5$. $T_{ij}^{(n)}$. The invariants $\lambda_m$ are functions of non-dimensional $\tilde S$ and $\tilde \Omega$ which are normalized by time scale $1/\omega$. 

For 2-D flows, we only consider the first three base tensors and first two nonzero invariants, which are given by:
\begin{equation}
\begin{aligned}
    T_{ij}^{(1)} &= \tilde{S}_{ij} & \lambda_1 &= \tilde{S}_{mn}\tilde{S}_{nm}, \\
    T_{ij}^{(2)} &= \tilde{S}_{ik}\tilde{\Omega}_{kj} - \tilde{\Omega}_{ik}\tilde{S}_{kj} & \lambda_2 &= \tilde{\Omega}_{mn}\tilde{\Omega}_{nm}, \\
    T_{ij}^{(3)} &= \tilde{S}_{ik}\tilde{S}_{kj} - \frac{1}{3} \delta_{ij} \tilde{S}_{mn}\tilde{S}_{nm}.
\end{aligned}
\end{equation}

Apart from the correction on RSM anisotropy, an additional correction $R$ is imposed on the production term in the \komegasst{} transport equations to correct the TKE equilibrium assumption. Following the work in~\citep{Schmelzer2019,BENHASSANSAIDI2022}, the correction $R$ takes the form
\begin{align}  
    R = 2 k b^R_{ij} \frac{\partial U_i}{\partial x_j},
\end{align}
and the correction tensor $b^R_{ij}$ is modeled in the same way of $b^\Delta_{ij}$ as follows:
\begin{align}
    b^R_{ij} = \sum_{n=1}^{10} T_{ij}^{(n)} \alpha_n^R (\lambda_1,\dots,\lambda_5).
\end{align} 

In the following, we developed an LLMs-based symbolic regression method to discover the algebraic expressions of $b^\Delta_{ij}$ and $b^R_{ij}$ to correct the turbulence model using high-fidelity data. 

\section{Proposed method: \ourmethod{}}

This work targets using LLMs to automatically design the algebraic correction terms for the turbulence model in the RANS solver to make it a better approximation of expensive high-fidelity experimental results. The proposed framework using LLMs, named \ourmethod{}, is given in this section. To guarantee data consistency and avoid numerical stiffness, both the direct observation of Reynolds stress and the indirect output of the CFD model are involved in training the models. For the numerical robustness of CFD simulation, the functional complexity and convergence residual are considered in the optimization problem.

\subsection{AutoTurb framework}

\ourmethod{} utilizes an evolutionary search framework in which each individual represents a unique turbulence model distinguished by a distinct algebraic correction term. These correction terms are generated by leveraging LLMs to craft expressions based on specified inputs and outputs. Each newly formulated algebraic correction term gives rise to a novel turbulence model, which is subsequently assessed within the RANS solver. The discrepancy between the simulation outcomes and experimental data serves as the basis for scoring (objective value) each turbulence model. \ourmethod{} not only markedly reduces human labor but also perpetually fosters the exploration of innovative designs, capitalizing on the capabilities of LLMs combined with the evolutionary framework.

% consists of three blocks, as illustrated in Figure~\ref{fig:AutoTurb}, 1) prompt block, 2) LLMs-based search block,  and 3) simulation block. 
% The evolutionary search block iteratively guides LLMs to produce the best-fitness expression to recover the target field. It is the high-level block that iteratively calls the LLMs block for generation and the simulation block for evaluation. 
% The LLMs block is used to generate new algebraic expressions based on prompt strategies with some parent expressions selected from the population. 
% The evaluation block compiles a new turbulence model upon each new algebraic expression and performs a RANS simulation over the training set. The difference between the high-fidelity data and the simulation results from each new model is used as the fitness value for each model during the search process. 

\begin{figure}[htbp]
    \centering
    \includegraphics[width=1\linewidth]{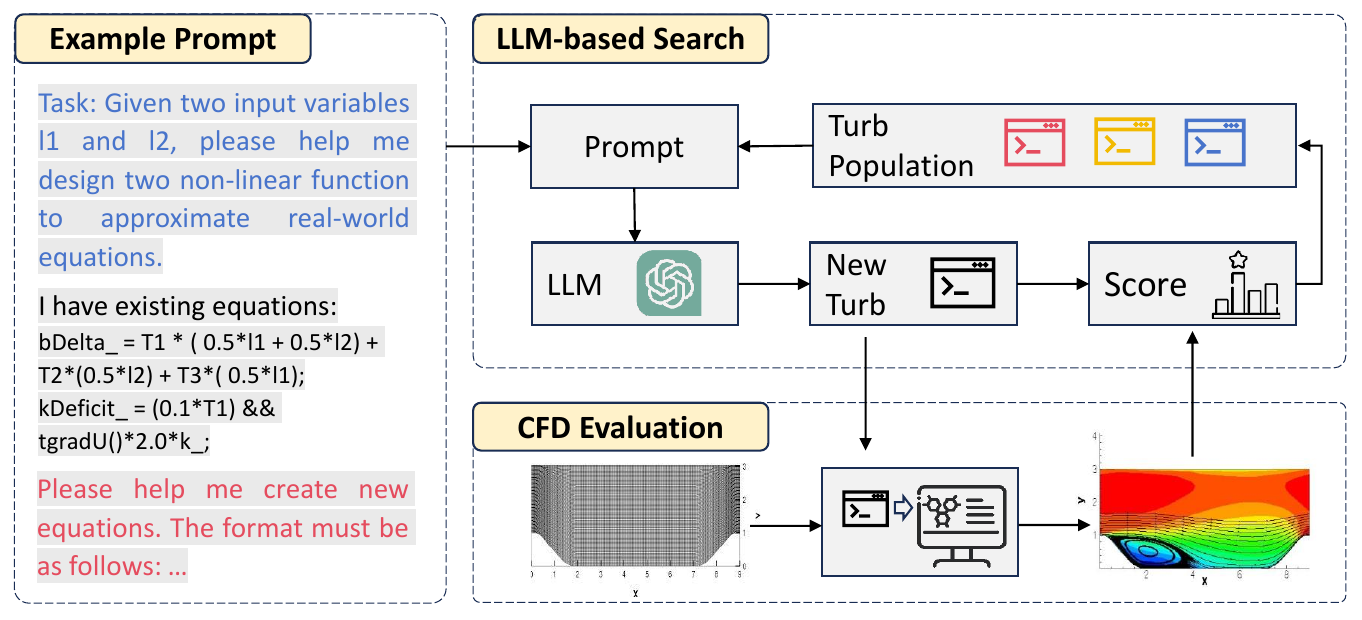}
    \caption{\ourmethod{} framework.}
    \label{fig:AutoTurb}
\end{figure}

\subsection{Objective formulation}

In this work, to maintain data consistency and avoid numerical instability, a CFD-driven approach is employed, with constraints placed on the complexity of the generated expressions. Both the internal Reynold stress and external mean velocity field by RANS simulation are utilized as the target quantity fields to train the models. Moreover, the vast design freedom provided by LLMs necessitates strict constraints on numerical convergence in CFD to avoid divergence or ill-conditioning and ensure a stable optimization process.

Two symbolic regression models for correcting the RSM and production term in \komegasst{} are trained simultaneously using the proposed algorithm.

Specifically, the following metrics are considered in the objective:

\begin{itemize}
    \item Minimizing the discrepancy between the target mean velocity field and the observed velocity field by running the RANS solver with corrective turbulence model:

    $f_1 = \left\| U^\star - U(b^\Delta_{ij}, R^\Delta_{ij})\right\|_2^2 $.
    
    \item Minimizing the discrepancy between the target TKE field and the observation $k$ by solving the turbulence transport equations:

    $f_2 = \left\| k^\star - k(b^\Delta_{ij}, R^\Delta_{ij})\right\|_2^2 $.
    
    \item Reducing the complexity of the expressions regarding the number of functional operators $n_o$. We assume that expressions with $n_o \leq 10$ are preferred, a segmented function is given by

    $f_3 = \left\{
    \begin{aligned}
        &\sqrt{n_{o}+1000}/\sqrt{1001},   \qquad &\text{if} \quad n_{o} \leq 10,\\
        &\sqrt{(n_{o})^2+1000-90}/\sqrt{1001}, \qquad &\text{if} \quad n_{o} > 10.
    \end{aligned}
    \right.$
    \begin{figure}[htbp]
        \centering
        \includegraphics[width=0.4\linewidth]{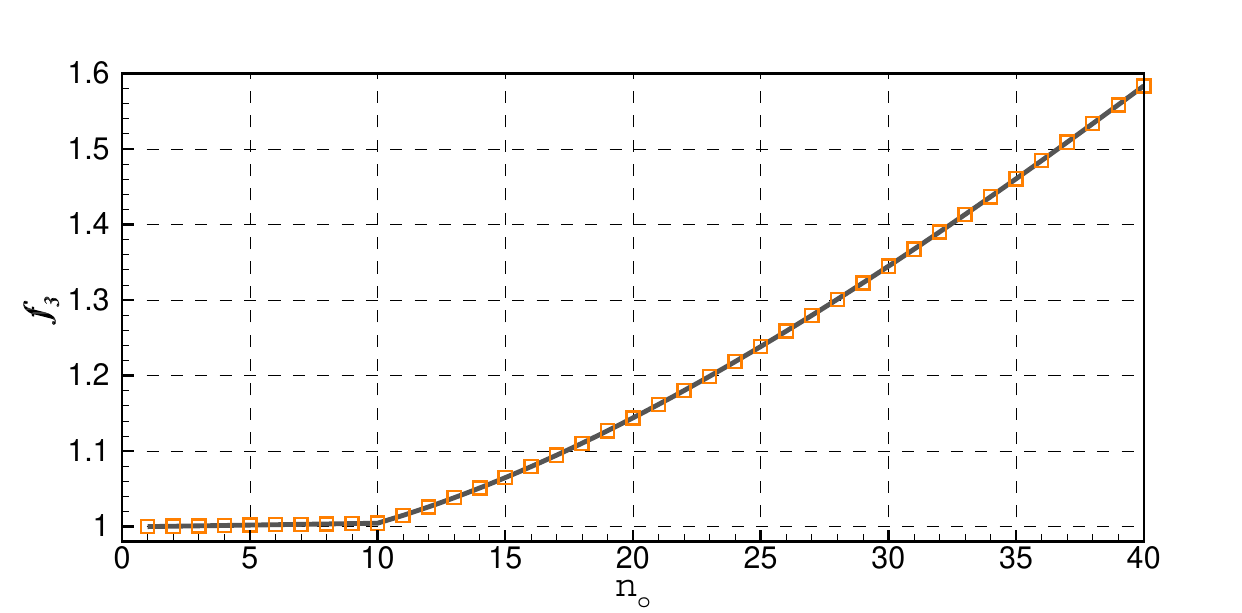}
        \caption{variation of $f_3$ versus the number of operators.}
    \end{figure}

    \item Ensuring the convergence of simulation by restricting the residual of the resolved mean pressure $Res(P)$ below $10^{-6}$:

    $f_4 = \left\{
    \begin{aligned}
        &1, \qquad           &\text{if} \quad Res \leq 10^{-6},\\
        &Res/10^{-6}, \qquad &\text{if} \quad Res > 10^{-6}.
    \end{aligned}
    \right.$
\end{itemize}

\noindent The first two metrics are computed and averaged over all mesh points of the training data set. The superscript $\star$ denotes quantities evaluated from the high-fidelity data. The last two merits or constraints are used to reduce model complexity and avoid over-fitting and divergence, as LLMs have the capability of producing expressions of arbitrary form and length. The final objective is as follows:

\begin{equation}
    f = (f_1 + 0.5*f_2) * f_3 * f_4.
\end{equation}

\subsection{Search method}

We adopt an evolutionary algorithm to iteratively prompt LLMs to generate new expressions to continuously explore novel and elite designs. We maintain a population of $N$ pairs of expressions, denoted as $P = \{e_1, \dots, e_N\}$, at each generation. Each pair of expression $e_i$ corresponds to a distinct turbulence model. The model will undergo evaluation in the RANS solver and is assigned an objective value $f(e_i)$. 

In the initialization, we let LLMs generate a population of expression pairs using the \textit{Initialization Prompt} illustrated in~\ref{app3}. During evolution, four \textit{Evolution Prompt} strategies with diverse tradeoffs on exploration and exploitation are adopted to generate new algebraic expressions. The details of prompt strategies are introduced in~\ref{app3}. In each generation, each strategy is executed $N$ times to produce $N$ new pairs of expressions. These newly generated expressions of two correction terms are evaluated in the RANS solver. A maximum of $4N$ pairs are added to the population in each generation. Subsequently, the top $N$ individuals with the best objective value from the current population are selected to form the population for the next generation.

The search method is outlined as follows:

\textbf{Step 0 Initialization:} The population $P$ of $N$ pairs of expressions $\{e_1, \ldots, e_N\}$ is initialized by prompting LLMs using the \textit{Initialization Prompt}.

\textbf{Step 1 LLM-based Model Search:} If the stopping condition is not met, four \textit{Evolution Prompt} strategies are simultaneously employed to generate $4N$ new expression pairs. For each prompt strategy, the following process is repeated $N$ times: \begin{itemize} \item Step 1.1: Select parent expressions from the current population to construct a prompt for the strategy. \item Step 1.2: Request LLM to generate a new design of expression pair along with its corresponding code implementation. \item Step 1.3: Evaluate the resulting new turbulence model in RANS solver to determine its objective value. \item Step 1.4: Add the new model to the current population if both the model and code are feasible. \end{itemize}

\textbf{Step 2 Population Management:} The top $N$ individuals in terms of objective value from the current population are selected to form the population for the next generation. The process then returns to \textbf{Step 1}.

% \begin{algorithm}[t]
%     \caption{Evolutionary Search for AutoTurb}
%     \label{alg:framework}
%     \KwIn{The number of population: $N_g$; Population size $N$; A given LLM.}
%     \KwOut{Best model $e^*$.}
%     \textbf{Initialization:} 

%     \For{$j=1,\dots,N$}{
%         \textbf{Model Creation:} create new individual $e_j$ given the target problem using LLM;
%         Evaluate $e_j$ and get fitness value $f(e_j)$;
%     }
    
%     Construct initial population $P=\{e_1,\dots,e_N\}$;
    
%     \For{$i=1,\dots,N_g$}{
%         \For {$j=1,\dots,N$}{
%             \textbf{Selection:} select a subset of input individuals $p_j=\{m_1,\dots,m_l \}$;

%             \textbf{Search Strategy} with probability $\theta_1$: create individual $o_j=\{m_1,...,m_s\}$ using LLM given the target problem and input subset $p_j$;

%         }
            
%         \textbf{Population management:} $P=P\cup \{o_1,...,o_N\}$, manage population $P$ to reduce the size from $(s+1)\dot N$ to $N$.
%         }
% \end{algorithm}

%%\subsection{Comparison with other symbolic regression methods \red{move to introduction?}}

%%The proposed method for symbolic regression using LLMs is compared with other representative methods, GEP, SpaRTA, FISR. We compare their common features in Table \ref{tab: methods}, although variants of these methods have been developed.

%

%
% \fei{

% It may consist of the following subsections
% \subsection{\ourmethod{} Framework}
% \subsection{Equation generation using large language models}
% \subsection{name of evaluation block}
% \subsection{name of our constraint methods,e.g., sparse and conditional number}

% }

\section{Numerical experiments settings}

\subsection{Experimental settings of \ourmethod{}}

In our experiments, the pre-trained GPT-3.5-turbo LLM is used. We configure the model with a temperature setting of 1.0 and a top-p value of 1.0. During the search process, the number of generations in AutoTurb is set to 20, the population size is 20, and the four prompt strategies are used simultaneously, which results in 1,600 evaluations. In evaluation, the maximum number of RANS iterations is set to 10,000, and the maximum running time for each instance is 2,000 seconds. The entire framework and the implementations are implemented in Python (with Python wrapper for OpenFoam) and executed on a single CPU i7-9700. %\red{internet address？}

\subsection{Datasets}

The details of high-fidelity data sets for training and cross-validate models are presented in this section. We use the high-fidelity data of widely studied separated flows over periodic hills at $Re = 10,595$ to train the symbolic regression model in the AutoTurb framework. Then, the flows over PHs with other steepness ratios at $Re = 5,600$, a converging-diverging channel at $Re = 12,600$, and a curved backward-facing step at $Re = 13,700$ are used to demonstrate the performance of the trained model and compare it with models found by other SR methods.

\subsubsection{Train cases}

\begin{figure} [ht]
    \centering
    \includegraphics[width=0.6\textwidth]{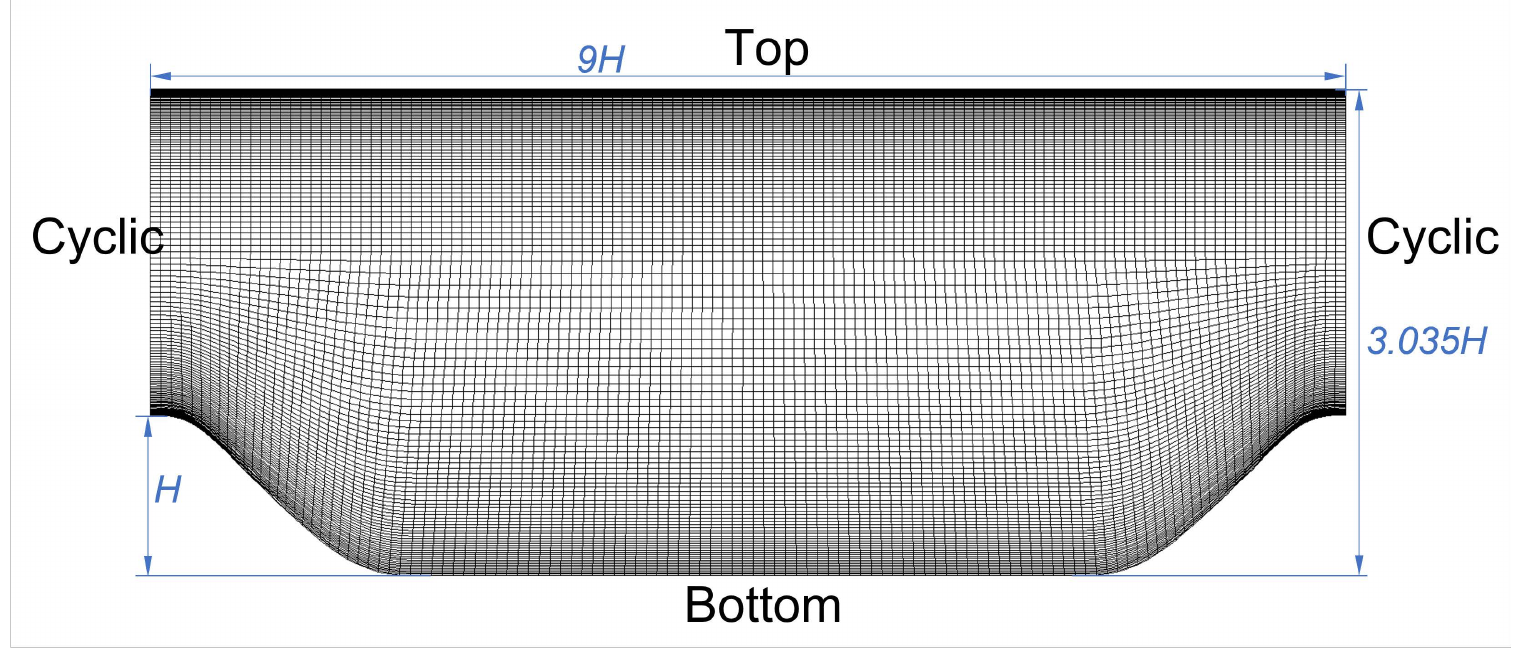}
    \caption{Geometry and computational grid for periodic hills (PH) case at $Re = 10595$.}
    \label{fig: PH10595grid}
\end{figure}

The highly-resolved LES data of periodic hills at $Re = 10595$ from ERCOFTAC Database Classic Collection case 81~\cite{FRÖHLICH2005} is used for training the model, which is shortened as ``PH". The height, spanwise width, and periodic streamwise length of the computational domain are $(9H, 3.035H, 4.5H)$ in terms of the hill height $H$, as shown in Fig.~\ref{fig: PH10595grid}. Cyclic boundary conditions are imposed at the hill crests. The top and bottom walls are treated as no slip. A constant bulk velocity is maintained at the inlet by adding a pressure gradient source term. The Reynolds number $10595$ is based on the bulk velocity and the hill height. 

The RANS simulation utilizes a computational grid with $120 \times 130$ cells in the stream-wise and wall-normal directions for spatial discretization. The simulation is conducted using the open-source finite-volume solver OpenFOAM. Referred to~\citep{McConkey2021}, the same numerical schemes are consistently applied across all RANS simulations in this study. The SimpleFoam solver, which employs the semi-implicit method for pressure-linked equations-consistent (SIMPLEC) algorithm, is used to simulate the 2-D incompressible, viscous, and steady flows.  For turbulence modeling, the~\komegasst{} model is adopted as the baseline turbulence closure model. 
The numerical schemes used include a second-order upwind scheme for the convective terms in the momentum equations. A first-order upwind scheme for the convective terms in the turbulence transport equations. A second-order central difference scheme for the diffusive terms.
The generalized geometric algebraic multigrid (GAMG) solver and the preconditioned bi-conjugate gradient (PBiCG) solver are used to solve the pressure equation and all other equations.

\subsubsection{Cross-validation cases}\label{sec:cvcases}
A set of separated flows with varying geometries and Reynolds numbers are used to validate the generalization capability of the trained models, as shown in Table~\ref{tab: cvcases}.
\begin{small}
\begin{table} [ht]
    \centering
    \caption{Cross-validation test cases.}
    \label{tab: cvcases}  
    \begin{tabular}{ccccc}
        \toprule[1pt]
        Flow cases & Notation & Re & No. of grids & Ref  \\
        \hline 
        Periodic hills with different steepness & PH$\alpha$ & 5600 & 99 $\times$ 149 & \citep{XIAO2020104431} \\
        Converging-diverging channel & CD & 12600 & 140 $\times$ 150 & \citep{Laval2011} \\
        Curved backward-facing step & CBF & 13700 & 140 $\times$ 180 & \citep{Bentaleb2020} \\
        \bottomrule[1pt]
    \end{tabular}
\end{table}    
\end{small}

\begin{enumerate}

\item [1)] Periodic hills. 
Four cases of periodic hills with steepness ratios $\{\alpha = 0.8, 1.0, 1.2, 1.5\}$ at $Re = 5,600$ are used for validation. For each case, a computational grid consisting of $99 \times 149$ cells in stream-wise and wall-vertical directions is employed for RANS simulation. An illustration of the case with steepness $\alpha = 0.8$ is shown in Fig.~\ref{fig:PH5600grid}. DNS simulation results are provided by Xiao et. al.~\citep{XIAO2020104431} using the Incompact3d solver and interpolated to RANS grids for comparison. It is shown that the flow separates after the first hill crest.
As the hills become flatter, the separation region decreases in size.
The separation bubbles are over-estimated by the baseline \komegasst{} closure model. The case with $\alpha = 0.5$ is not included in this work, as the baseline model can provide simulation results with acceptable accuracy for this configuration. The remaining cases are abbreviated as ``PH$\alpha_{0.8}$", ``PH$\alpha_{1.0}$", ``PH$\alpha_{1.2}$", ``PH$\alpha_{1.5}$".
\begin{figure} [ht]
    \centering
    \includegraphics[width=0.6\textwidth]{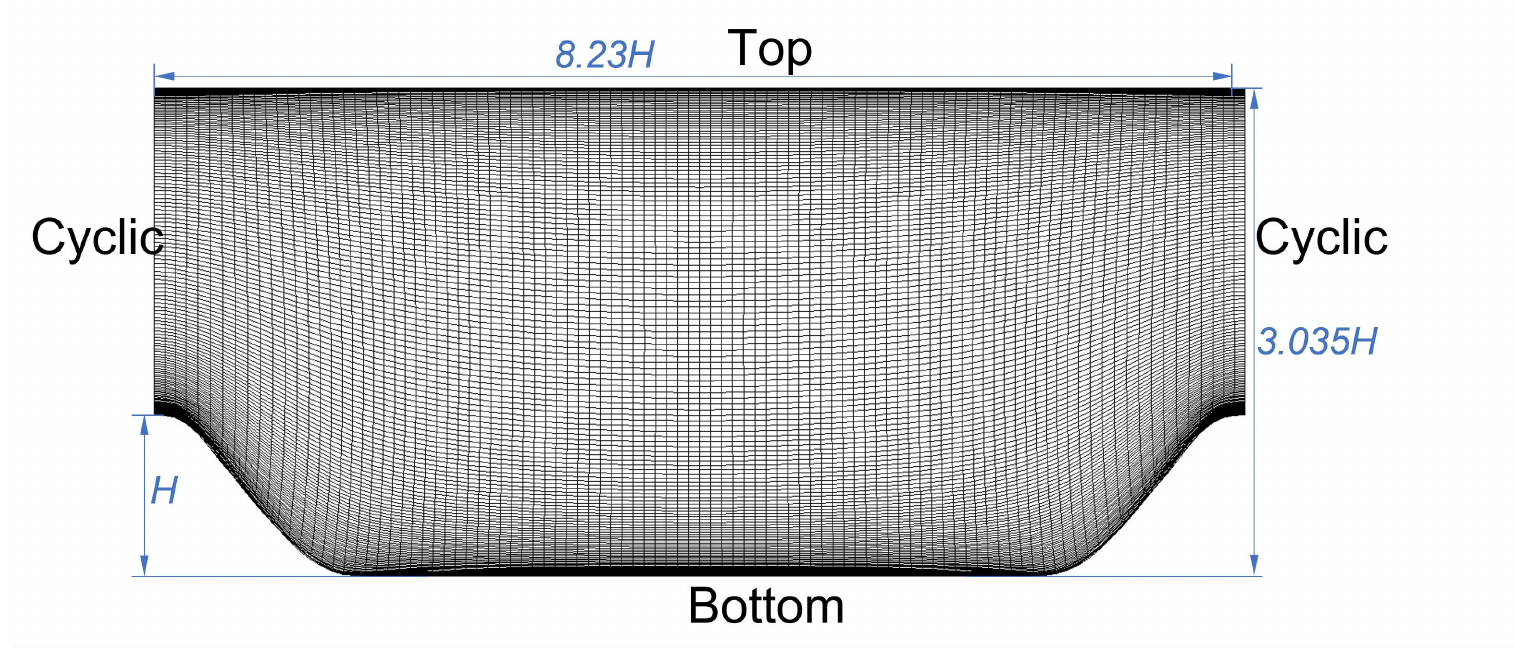}
    \caption{Geometry and computational grid for periodic hills case with $\alpha = 0.8$ (PH$\alpha_{0.8}$), $Re = 5,600$.}
    \label{fig:PH5600grid}
\end{figure}

\item [2)] Converging-diverging channel. 
The second case, labeled as ``CD", is a channel flow with an asymmetric hill that has a height of $2/3H$, located near $x = 5.21564$. The Reynolds number for this flow, based on the channel half-height and the maximum velocity $U_{max}$ at the inlet, is 12600. A fully developed channel flow at the same Reynolds number is applied as the inlet condition. High-fidelity DNS data from Laval et al.~\citep{Laval2011} indicate the presence of a small separation region on the lee-side of the hill crest. For the RANS simulations, a computational grid with 140 x 150 cells is used. The computational grid and geometry for this case are shown in Fig.\ref{fig:CD12600grid}.
\begin{figure} [ht]
    \centering
    \includegraphics[width=0.6\textwidth]{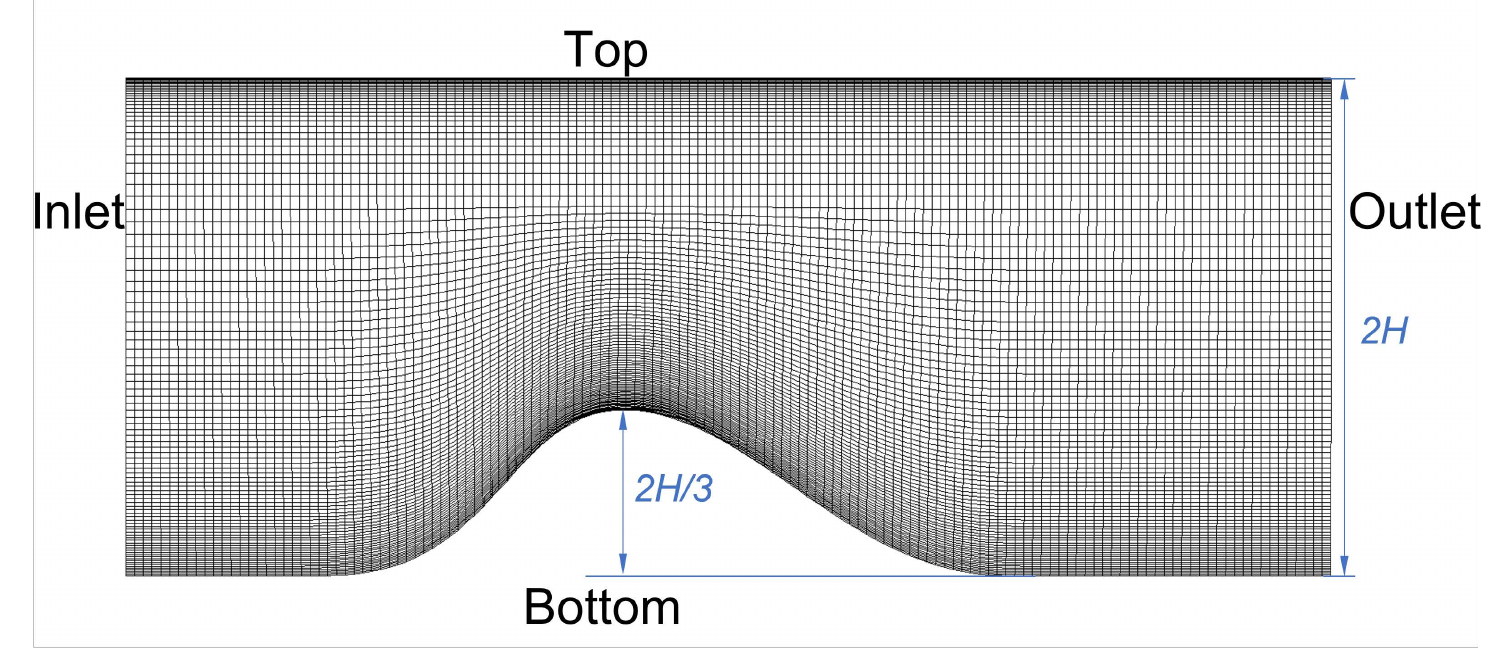}
    \caption{Geometry and computational grid for converging and diverging channel (CD) case, $Re = 12,600$.}
    \label{fig:CD12600grid}
\end{figure}

\item [3)] Curved backward-facing step.
The third test case, labeled as ``CBF", involves a 2D flow over a gently curved backward-facing step with a height $H$. The upstream channel height is $8.52H$ and the Reynolds number, based on $U_{max}$ at the inlet and step height, is 13700. A fully developed channel flow at the same Re serves as the inlet condition. High-fidelity LES data from Bentaleb et al.~\citep{Bentaleb2020} are used for training the model. The RANS simulations are performed on a computational grid of 140 x 180 cells, depicted in Fig. \ref{fig:CBF13700grid}. 
\begin{figure} [htbp]
    \centering
    \includegraphics[width=0.6\textwidth]{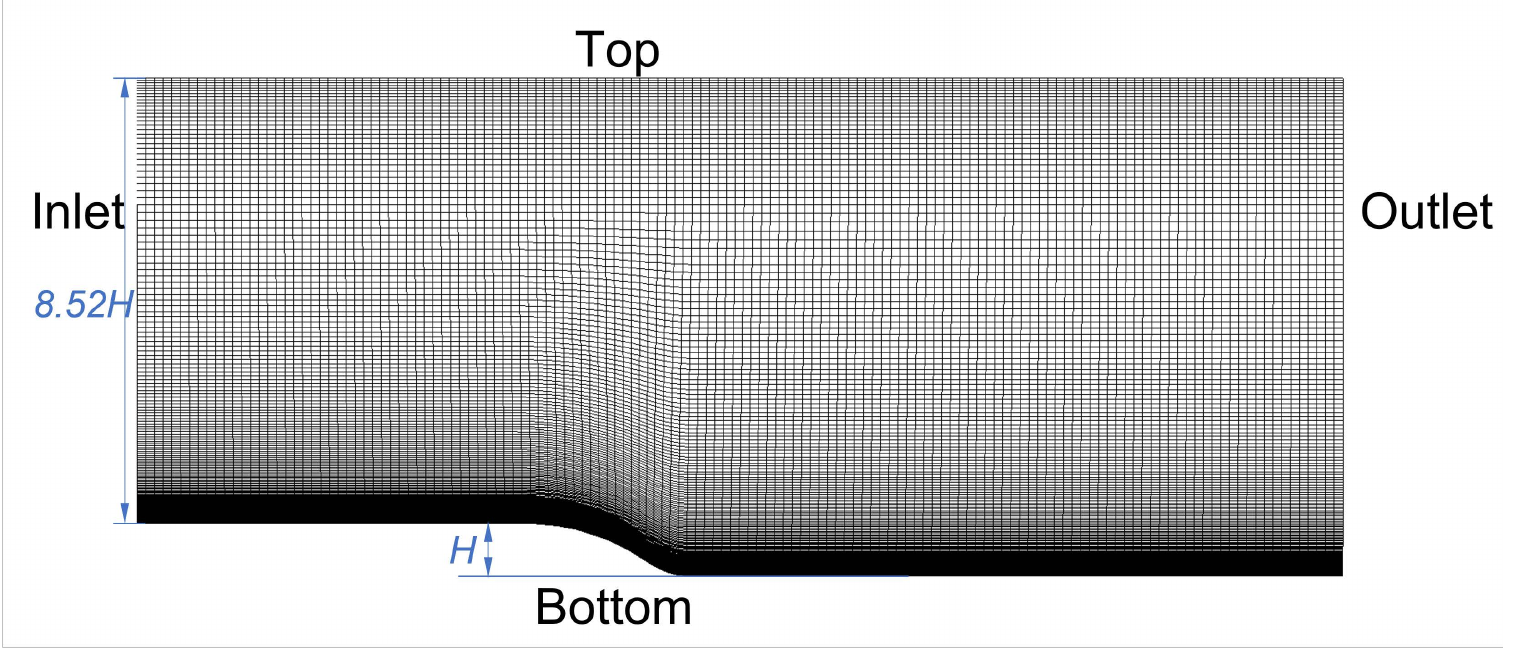}
    \caption{Geometry and computational grid for curved backward-facing step (CBF) case, $Re = 13,700$.}
    \label{fig:CBF13700grid}
\end{figure}

\end{enumerate}

\section{Results}

\subsection{Model discovery}

\begin{figure} [htbp]
    \centering
    \includegraphics[width=0.6\textwidth]{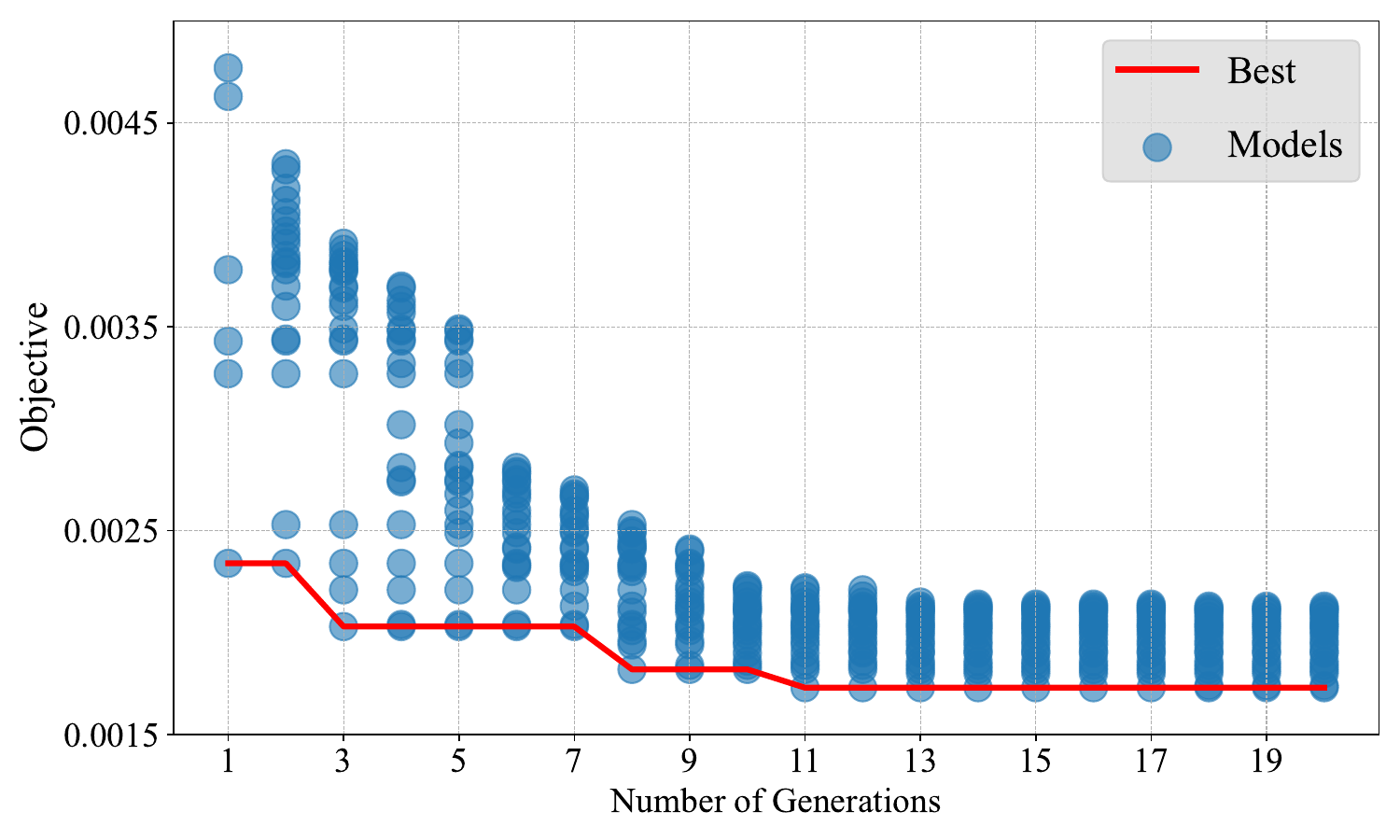}
    \caption{Convergence curves of objective function (Each blue sample formulates a turbulence model).}
    \label{fig:converge}
\end{figure}

The convergence curves for the LLM-based evolutionary search are depicted in Fig.~\ref{fig:converge}. The y-axis represents the hybrid objective value, while the x-axis denotes the generation number. Each sample corresponds to a turbulence model designed by LLMs during the evolutionary process. The red line highlights the best-performing model in each population as the evolution progresses. A clear convergence is observed within 20 generations, with the best objective value decreasing from 0.0023 to 0.0017. Initially, two models in the population failed to converge in the simulation, exhibiting a large residual error (these are not shown in the convergence curve due to their extremely high objective values). In contrast, after 20 generations of evolution, all models in the final population demonstrate reduced complexity and enhanced robustness during simulations.

The optimum algebraic expressions for the two corrective terms found by \ourmethod{} are 
\begin{equation}\label{ourmodel}
    \text{Model-LLMs:} \left\{
    \begin{aligned}
        &\Delta b_{ij} = 0, \\
        &\Delta R_{ij} = 2k\partial_j U_i [(sin(l1) + 0.5) \times T_{ij}^1 + T_{ij}^2. ]
    \end{aligned}
    \right.
\end{equation}

\begin{figure*} [htbp]
    \centering
    \subfigure[High-fidelity data]
    {\includegraphics[width=0.5\textwidth]{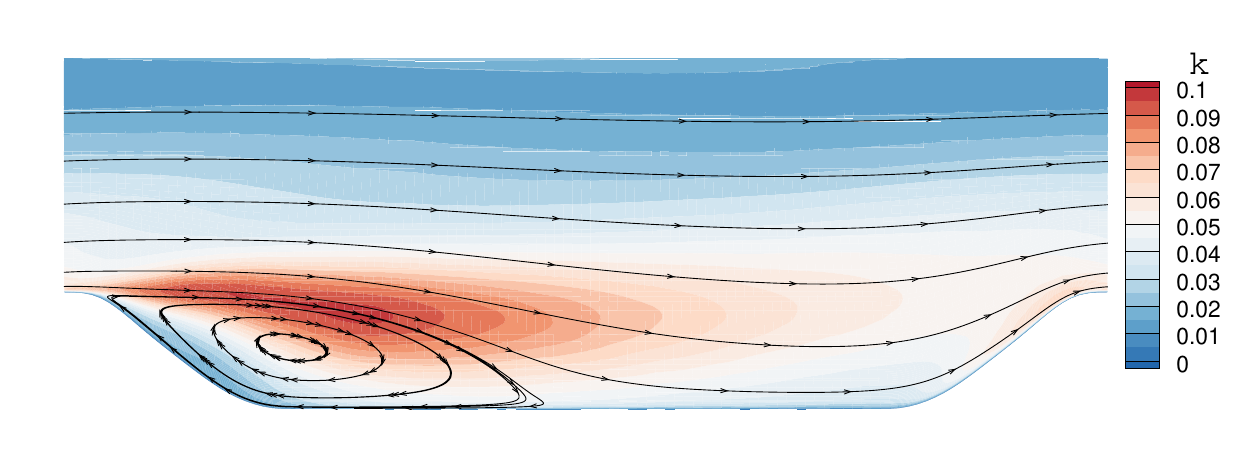}    }
    \\
    \begin{minipage}{0.48\textwidth}
        \centering
        \subfigure[Baseline model \komegasst{}]
        {\includegraphics[width=1\textwidth]{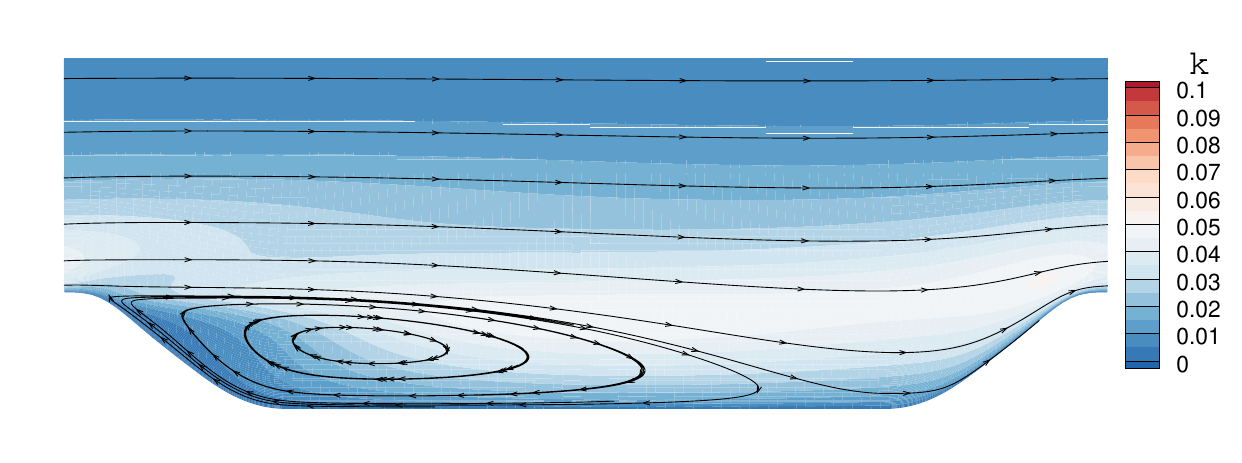}    }
    \end{minipage}
    \hfill
    \begin{minipage}{0.48\textwidth}
        \centering
        \subfigure[Model-LLMs]
        {\includegraphics[width=1\textwidth]{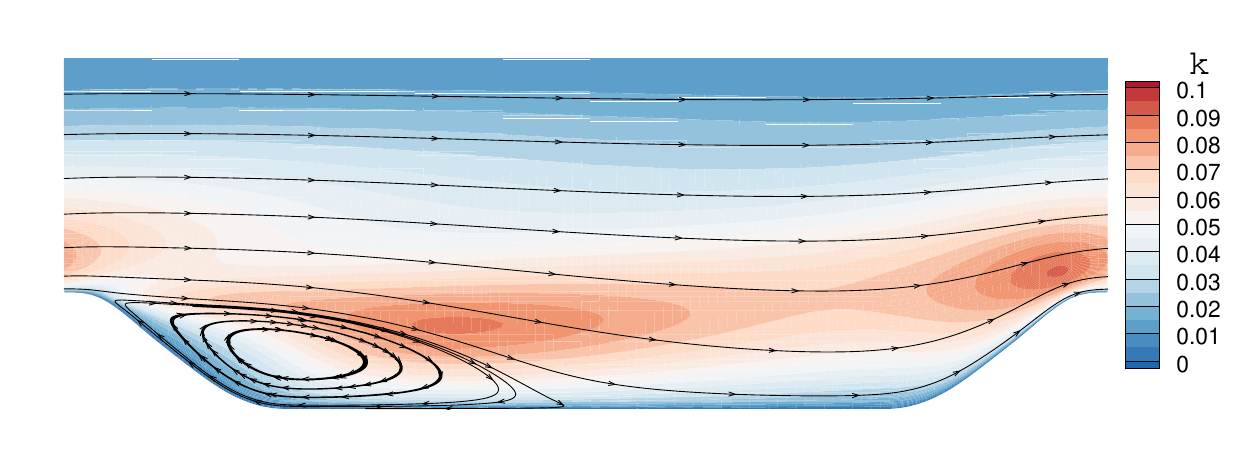}    }
    \end{minipage}
    \caption{Contours of TKE $k$ with velocity streamlines.  }
    \label{fig:objcompare}
\end{figure*}

\begin{figure} [htbp]
    \centering
    \includegraphics[width=0.55\textwidth]{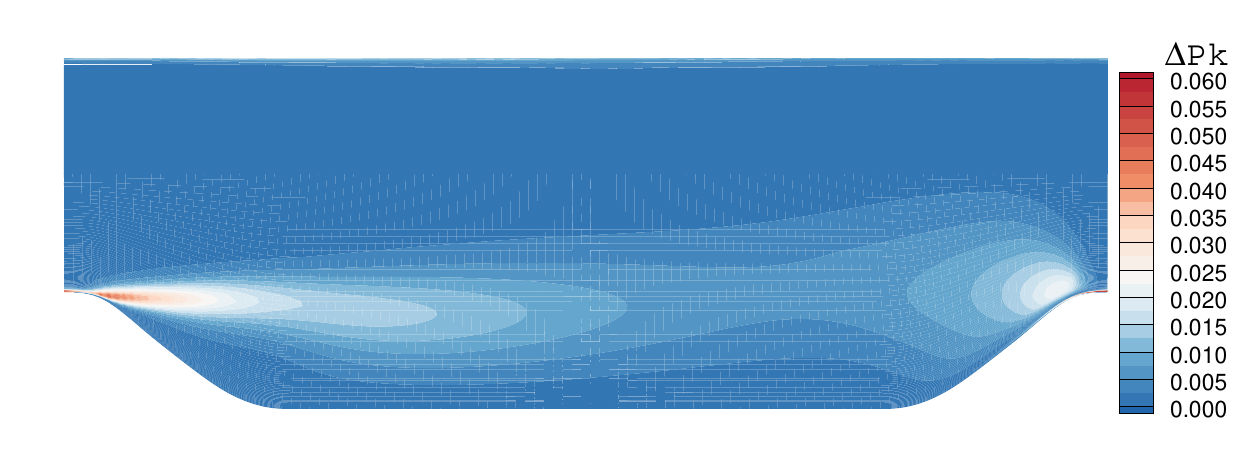}
    \caption{Correction to the production term $P_k$ by Model-LLMs.}
    \label{fig:Pk}
\end{figure}

\begin{figure*} [htbp]
    \centering
    \begin{minipage}{0.48\textwidth}
        \centering
        \subfigure[Baseline model \komegasst{}]
        {\includegraphics[width=1\textwidth]{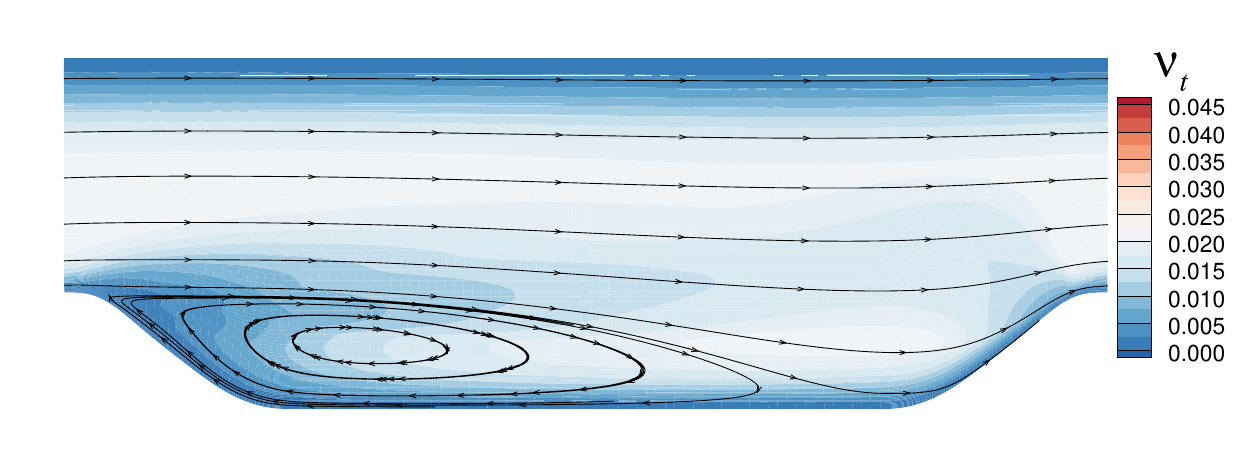}    }
    \end{minipage}
    \hfill
    \begin{minipage}{0.48\textwidth}
        \centering
        \subfigure[Model-LLMs]
        {\includegraphics[width=1\textwidth]{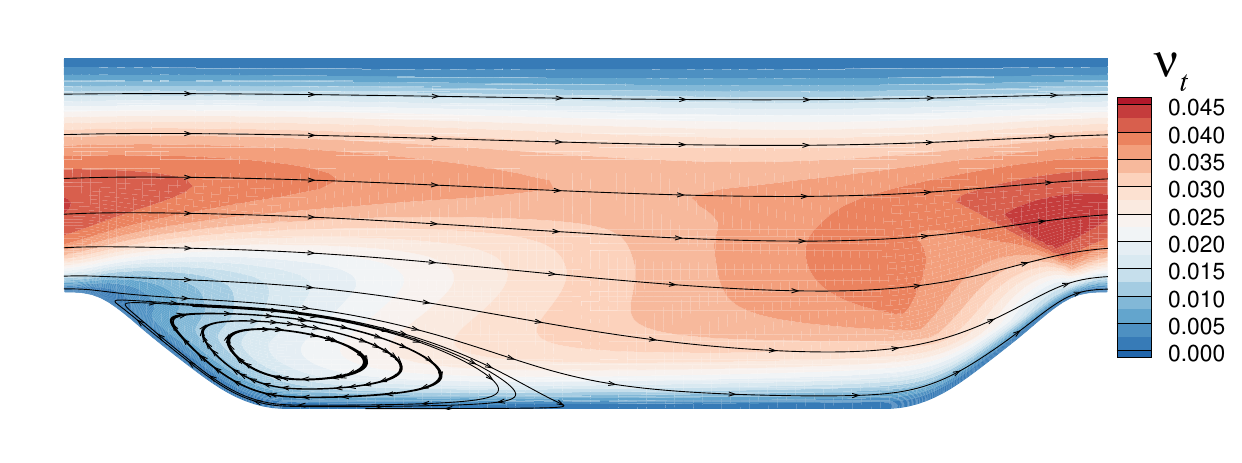}    }
    \end{minipage}
    \caption{Contours of eddy viscosity with velocity streamlines.  }
    \label{fig:nut}
\end{figure*}

\noindent The resulting expressions maintain their elegant simplicity due to the complexity and convergence constraints imposed on the objective. This simplicity is crucial in avoiding overfitting and ensures robust performance across different scenarios.
The observed velocity streamlines and TKE $k$ fields using the discovered model is shown in Fig. \ref{fig:objcompare}. Both the velocity and $k$ fields are significantly improved.
In this particular training case, the linear eddy-viscosity constitutive model is preserved, as the correction term $\Delta b_{ij}$ remains zero. However, a correction term is introduced to the production $P_k$ term in the TKE transport equation, as depicted in Fig. \ref{fig:Pk}.
This adjustment significantly amplifies the magnitude of $P_k$, leading to a higher value of $k$. 
The modification of $P_k$ increases both turbulence intensity and eddy viscosity $\nu_t$, which enhances the model’s resistance to flow separation. As shown in Fig. \ref{fig:nut}, with the increase of eddy viscosity, a smaller separation bubble is observed. 

In the following section, we systematically evaluate the performance of our model by incorporating it into the RANS equations to solve flow fields, and then compare the results with high-fidelity data. 
In addition, models identified by other SR methods are integrated into the RANS solver for comparison. Although plenty of SR methods have been developed, only a few models with explicit algebraic expressions of similar corrective terms for separated flows are available. 
Two models from Benhassan-Saidi et al. \citep{BENHASSANSAIDI2022}, developed using the CFD-driven SpaRTA method, are included in the comparison. ``Model 1'' was trained against high-fidelity Reynolds stress data, and ``Model 2'' was trained using indirect data including mean velocity and skin friction. 
A ``Model frozen'' reported by Schmelzer et al. \citep{Schmelzer2019}, which was trained using k-corrective frozen-RANS with the SpaRTA method, is also investigated. These models are described in detail below:
\begin{align}
    &\text{Model 1:} \left\{
    \begin{aligned}\label{model1}
        &\Delta b_{ij} = \left(-0.147 I_1^2 \right) \times T_{ij}^1 + \left(- 0.26791\right) \times T_{ij}^2,\\
        &\Delta R_{ij} = 2k\partial_j U_i(-0.46018 \times T_{ij}^1 - 0.16779 \times T_{ij}^3).
    \end{aligned}
    \right. \\
    &\text{Model 2:} \left\{
    \begin{aligned}\label{model2}
        &\Delta b_{ij} = \left(-0.28356\right) \times T_{ij}^1 + \left(- 0.14738 I_2^2\right) \times T_{ij}^2,\\
        &\Delta R_{ij} = 2k\partial_j U_i(-0.10375 \times I_2  - 0.28833 \times I_1 I_2) T_{ij}^3. 
    \end{aligned}
    \right. \\
    &\text{Model-frozen:} \left\{
    \begin{aligned}\label{modelfrozen}
        &\Delta b_{ij} = 0,\\
        &\Delta R_{ij} = 2k\partial_j U_i(-0.39 \times T_{ij}^1).
    \end{aligned}
    \right.
\end{align}

\subsection{Model evaluations}

\begin{figure} [htbp]
    \centering
    \subfigure[streamwise velocity profiles ($U_1/U_b + x/H$)]
    {\includegraphics[width=0.6\textwidth]{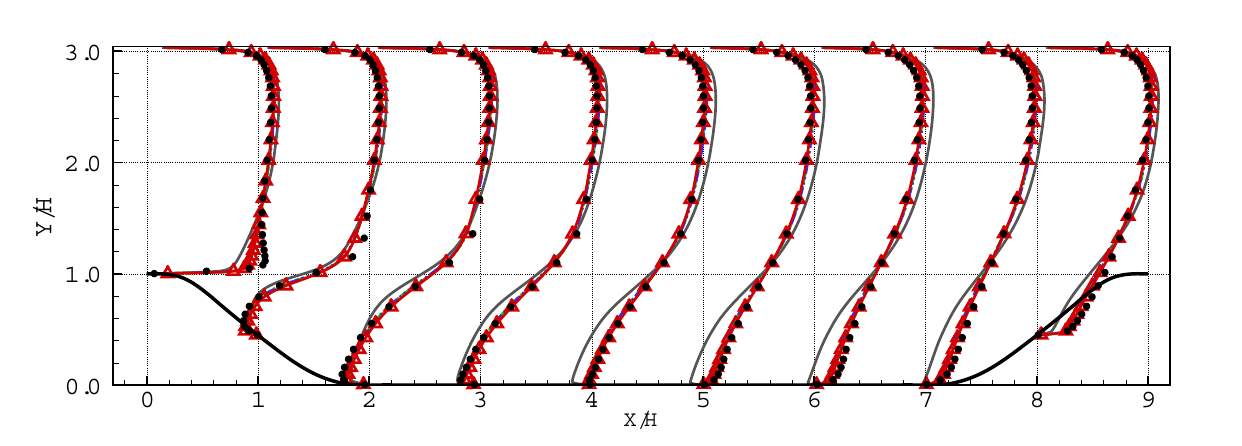}    }
    \\
    \subfigure[turbulent kinetic energy ($6k/U_b^2 + x/H$)]
    {\includegraphics[width=0.6\textwidth]{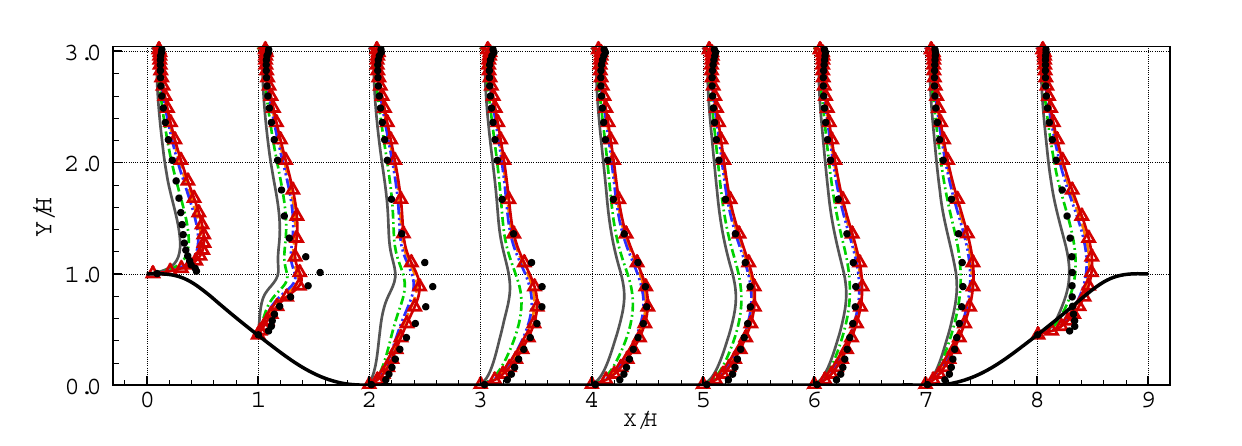}    }
    \\
    \subfigure[Reynolds shear stress ($20\tau_{12}/U_b^2 + x/H$)]
    {\includegraphics[width=0.6\textwidth]{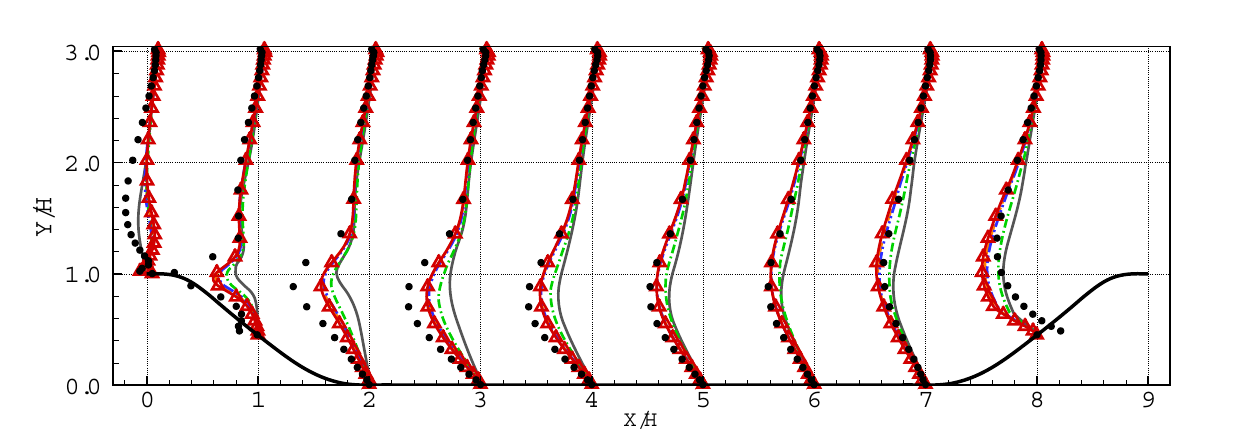}    }
    \caption{Estimation for various SR models for flow over Periodic hills $Re = 10,595$. 
    High-fidelity data: ({\color{black} $\cdot \cdot \cdot$}), 
    \komegasst{}: ({\color{black}{\textbf{\textemdash}}}), 
    Model 1: ({\color{orange}- - - -}), 
    Model 2: ({\color{green}-$\cdot$-$\cdot$-}), 
    Model-frozen: ({\color{blue}-$\cdot$$\cdot$-$\cdot$$\cdot$-}), 
    Model-LLMs: (\trianglewithline). }
    \label{fig:PH10595}
\end{figure}

The four models, Model-LLMs, Model 1, Model 2, and Model-frozen, are all trained by the high-fidelity data from flow over periodic hills at $Re = 10,595$. Mean squared errors (MSE) of the predictions for streamwise mean velocity field $U_1$, turbulent kinetic energy $k$, and Reynolds shear stress anisotropy $\tau_{12}$ are shown in Table~\ref{tab: msePH}. The profiles at $x/H = 0.05$, $1$, $2$, $3$, $4$, $5$, $6$, $7$, $8$ are shown in Fig.~\ref{fig:PH10595}. 
Compared with the baseline model, the mean velocity field is greatly improved using the corrected turbulence closure. A smaller separation region is captured, aligning well with the high-fidelity LES data. 
Meanwhile, the discovered model provides a more accurate prediction for the Reynolds stress, which is proven by the improved prediction for turbulent kinetic energy $k$, and Reynolds shear stress $\tau_{12}$ in Fig.~\ref{fig:PH10595} (b) and (c). The MSE of mean velocity and $k$ is reduced by $14.2\%$ and $30.4\%$, respectively, compared to the baseline.

All four models greatly improve the simulation results compared with the baseline. Among them, the present model achieves the most accurate prediction for the velocity field, while comparable results for $k$ and $\tau_{12}$ are provided by Model-LLMs, Model 1, and Model-frozen. It is due to more weight being assigned to achieving an accurate velocity field compared to the TKE $k$. Model 2 performs worse, as it was trained using only indirect data. 
\begin{table} [htbp]
    \centering
    \caption{MSE of the prediction for training data using various SR models (PH $Re = 10,595$). }
    \label{tab: msePH}  
    \begin{tabular}{cccc}
        \toprule[1pt]
        models & $U_1$ & $k$ & $\tau_{12}$  \\
        \hline 
        \komegasst{}  & 4.39013e-3 & 3.90736e-4 & 3.69883e-5  \\
        Model-LLMs    & 6.22571e-4 & 1.18714e-4 & 2.05218e-5 \\
        Model 1       & 6.90418e-4 & 1.10777e-4 & 2.03590e-5 \\
        Model 2       & 7.64242e-4 & 1.88782e-4 & 2.68433e-5 \\
        Model-frozen  & 8.07702e-4 & 1.12319e-4 & 2.03887e-5 \\
        \bottomrule[1pt]
    \end{tabular}
\end{table}

\subsection{Model cross-validation}
\begin{figure} [htbp]
    \centering
    \subfigure[streamwise velocity profiles ($U_1$)]
    {\includegraphics[width=0.6\textwidth]{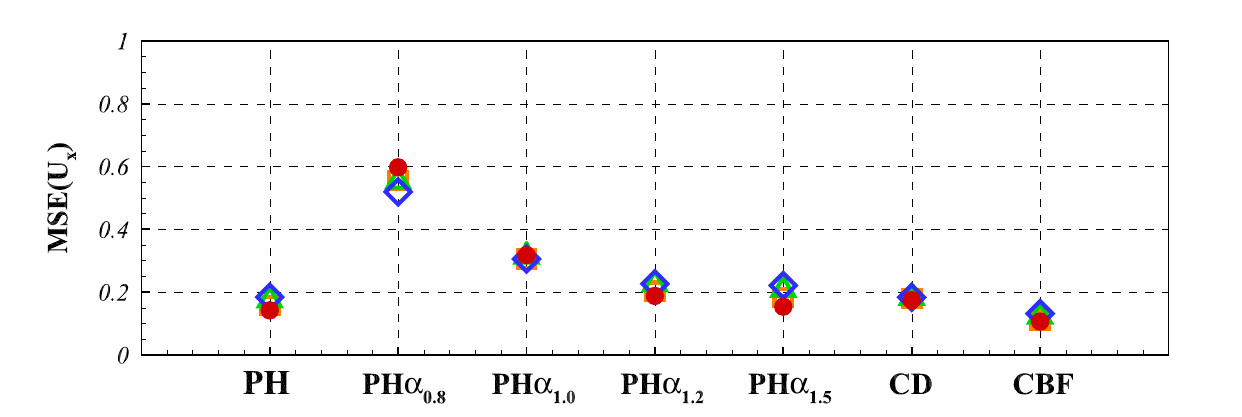}    }
    \\
    \subfigure[turbulent kinetic energy ($k$)]
    {\includegraphics[width=0.6\textwidth]{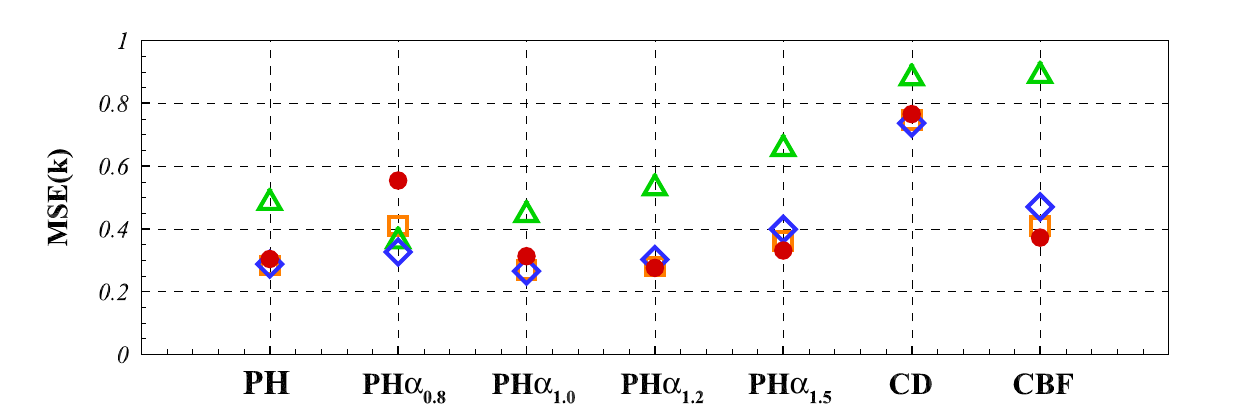}    }
    \\
    \subfigure[Reynolds shear stress ($\tau_{12}$)]
    {\includegraphics[width=0.6\textwidth]{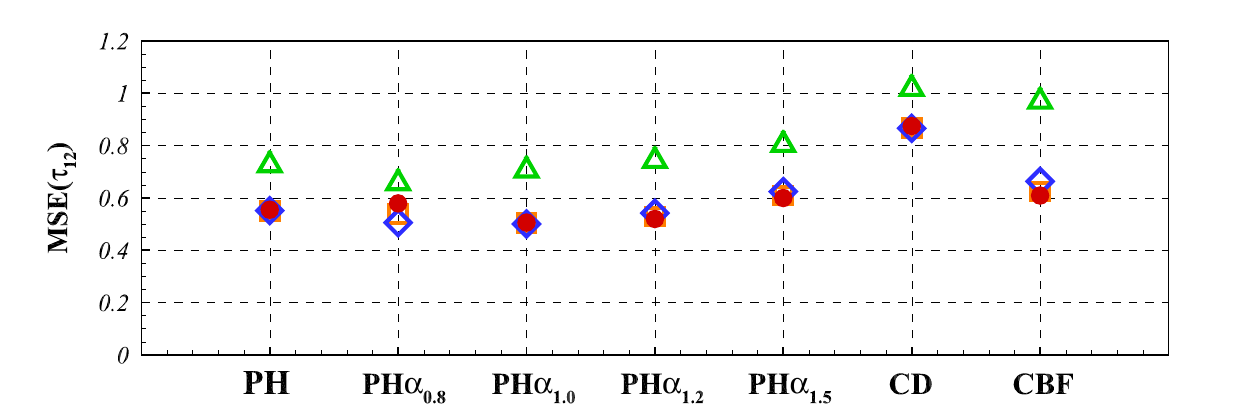}    }
    \caption{Mean squared errors for flow quantities of cross-validation cases using various models. 
    Model 1: (\orangesquare),
    Model 2: (\greentriangle),
    Model-frozen: (\bluediamond),
    Model-LLMs: (\redcircles). }
    \label{fig:mse}
\end{figure}

The discovered model is further validated on a set of separated flows over different geometry and Reynolds numbers, as given in Table \ref{tab: cvcases} in section \ref{sec:cvcases}. The MSE of the streamwise mean velocity field $U_1$, turbulent kinetic energy $k$, and Reynolds shear stress anisotropy $\tau_{12}$ by the models for all the test cases are depicted in Fig. \ref{fig:mse}. And the details are given in Tab. \ref{tab: mseall} in \ref{app1}. In all cases, the errors are normalized with the MSE of the baseline \komegasst{} model. The profiles of $U_1$, $k$, and $\tau_{12}$ at a sequence of streamwise locations are displayed in \ref{app2}.
All of the models significantly improve the simulation results, compared to the baseline model. Significantly improved velocity fields are retrieved with more accurate separation regions for all cases. The improvement of the velocity field is relatively less in the PH$\alpha_{0.8}$ case, as the error of the baseline model is already lower in this case. 

It is observed that the present model produces the smallest error for the velocity field in the training case PH, PH$\alpha_{1.2}$, and PH$\alpha_{1.5}$ at $Re = 5,600$, as well as in the CD, and CBF cases. Meanwhile, it provides the most accurate predictions for $k$ in the training case PH, PH$\alpha_{1.2}$, PH$\alpha_{1.5}$, and CBF, and generates the best predictions for $\tau_{12}$ in all cases except for the PH$\alpha_{0.8}$ case. Overall, the present model demonstrates the best performance in terms of $U_1$ and Reynold stress across most cases. 

The accuracy of the present model is Slightly better than Model 1 and Model-frozen. Model 1, which is trained solely using the Reynolds stress, performs best in predicting Reynolds stress anisotropy for most cases. However, the present model achieves lower errors in velocity prediction compared to Model 1, as it is involved in the objective function.
Model-frozen shows slightly worse performance than the present Model-LLMs, as it incorporates direct and indirect data in an implicit offline manner. On the contrary, Model 2, trained using only indirect data, can not retrieve the Reynolds stress field as accurately as other models, though it still outperforms the baseline model.

\section{Conclusion}

In this work, a novel framework, \ourmethod{}, is proposed, leveraging large language models (LLMs) to automatically discover algebraic expressions for correcting turbulence closure models. This approach introduces two algebraic correction terms: one for the Reynolds stress model and another for the production term in the \komegasst{} TKE transport equation. LLMs are utilized to prompt the optimization problem using natural language and autonomously generate algebraic expressions for the correction terms. The objective function is formulated by integrating direct observations of TKE $k$ and indirect velocity outputs from the CFD model to ensure data consistency. Additionally, constraints on functional complexity and numerical convergence of corrected RANS solver are imposed to prevent divergence or ill-conditioning, ensuring a stable optimization process.

The proposed method is performed for separated flow over periodic hills at $Re = 10,595$. For the discovered model, the linear eddy-viscosity constitutive model is preserved, while the production term  $P_k$ is enlarged. Consequently, the eddy viscosity $\nu_t$ is increased, which enhances the model’s resistance to flow separation, and a smaller separation bubble is observed. Compared with the baseline model, the discovered model provides more accurate predictions for both the velocity field and the Reynolds stress.

The generalizability of the discovered model is validated on a series of 2D turbulent separated flow configurations with varying Reynolds numbers and geometries. The present Model-LLMs exhibit the best overall performance in predicting both $U_1$ and Reynolds stress across most cases. It surpasses Model 1 in velocity prediction and outperforms Model 2 in predicting Reynolds stress quantities. Despite Model-frozen being trained with both direct and indirect data, it still falls short of the discovered model, which benefits from a CFD-driven training approach.

The proposed AutoTurb framework using LLMs provides a promising paradigm for using LLMs to improve turbulence modeling in specific classes of flows. Its success suggests the potential for extending this approach to a broader range of flow types and promoting the use of LLMs in various other fields of CFD.

\newpage
\appendix
\section{Mean squared error of predicted flow quantities by various models}
\label{app1}
\begin{table} [htb]
    \caption{Mean-squared errors of the predictions for cross-validation cases using various SR models.}
    \label{tab: mseall}  
    \centering
%\subtable[Periodic hills $Re = 5600, \alpha = 0.5$]{
%    \begin{tabular}{ccccc}
%        \toprule[1pt]
%        models & modelLLMs & model1 & model2 & modelfrozen  \\
%        \hline 
%        $U_1$       & 1.56804 & 1.49246 & 1.48990 & 1.37632 \\
%        $k$         & 2.19966 & 1.36556 & 0.34490 & 0.95630 \\
%        $\tau_{12}$ & 1.32724 & 1.01183 & 0.63190 & 0.82756 \\
%        \bottomrule[1pt]
%    \end{tabular}
%}
\scalebox{0.7}{
\subtable[Periodic hills $Re = 5,600, \alpha = 0.8$]{
    \begin{tabular}{ccccc}
        \toprule[1pt]
        models & Model-LLMs & Model 1 & Model 2 & Model-frozen  \\
        \hline 
        $U_1$       & 0.59732 & 0.55572 & 0.55236 & 0.51910 \\		
        $k$         & 0.55390 & 0.41003 & 0.36072 & 0.32621 \\		
        $\tau_{12}$ & 0.57899 & 0.53899 & 0.65663 & 0.50548 \\
        \bottomrule[1pt]
    \end{tabular}}
}
\scalebox{0.7}{
\subtable[Periodic hills $Re = 5,600, \alpha = 1.0$]{
    \begin{tabular}{ccccc}
        \toprule[1pt]
        models & Model-LLMs & Model 1 & Model 2 & Model-frozen  \\
        \hline 
        $U_1$       & 0.31749 & 0.30496 & 0.31416 & 0.30559 \\		
        $k$         & 0.31281 & 0.26816 & 0.44471 & 0.26529 \\		
        $\tau_{12}$ & 0.50574 & 0.50401 & 0.70544 & 0.50052 \\
        \bottomrule[1pt]
    \end{tabular}}
}
\scalebox{0.7}{
\subtable[Periodic hills $Re = 5,600, \alpha = 1.2$]{
    \begin{tabular}{ccccc}
        \toprule[1pt]
        models & Model-LLMs & Model 1 & Model 2 & Model-frozen  \\
        \hline 
        $U_1$       & 0.18771 & 0.20242 & 0.2254  & 0.22640 \\		
        $k$         & 0.27540 & 0.28038 & 0.52984 & 0.30224 \\		
        $\tau_{12}$ & 0.51952 & 0.52875 & 0.74205 & 0.54179 \\	
        \bottomrule[1pt]
    \end{tabular}}
}
\scalebox{0.7}{
\subtable[Periodic hills $Re = 5,600, \alpha = 1.5$]{
    \begin{tabular}{ccccc}
        \toprule[1pt]
        models & Model-LLMs & Model 1 & Model 2 & Model-frozen  \\
        \hline 
        $U_1$       & 0.15356 & 0.18288 & 0.20766 & 0.22088 \\		
        $k$         & 0.33078 & 0.36053 & 0.65510 & 0.39868 \\		
        $\tau_{12}$ & 0.59956 & 0.60911 & 0.80335 & 0.62424 \\	
        \bottomrule[1pt]
    \end{tabular}}
}
\scalebox{0.7}{
\subtable[converging-diverging channel $Re = 12,600$]{
    \begin{tabular}{ccccc}
        \toprule[1pt]
        models & Model-LLMs & Model 1 & Model 2 & Model-frozen  \\
        \hline 
        $U_1$       & 0.17462 & 0.17986 & 0.18274 & 0.18334 \\		
        $k$         & 0.76540 & 0.74575 & 0.88018 & 0.73650 \\		
        $\tau_{12}$ & 0.87447 & 0.86706 & 1.01743 & 0.86604 \\	
        \bottomrule[1pt]
    \end{tabular}}
}    
\scalebox{0.7}{
\subtable[curved back-facing step $Re = 13,700$]{    
    \begin{tabular}{ccccc}
        \toprule[1pt]
        models & Model-LLMs & Model 1 & Model 2 & Model-frozen  \\
        \hline 
        $U_1$       & 0.10612 & 0.11126 & 0.12271 & 0.13153 \\
        $k$         & 0.37198 & 0.40803 & 0.88775 & 0.46985 \\
        $\tau_{12}$ & 0.60913 & 0.62428 & 0.96893 & 0.66300 \\
        \bottomrule[1pt]
    \end{tabular}}
}
\end{table}

\newpage
\section{Predicted flow fields by various SR models for cross-validation flows}
\label{app2}

\begin{figure*} [htbp]
    \centering
    \begin{minipage}{0.48\textwidth}
        \centering
        \subfigure[$20 U_1/U_b + x/H$]
        {\includegraphics[width=1\textwidth]{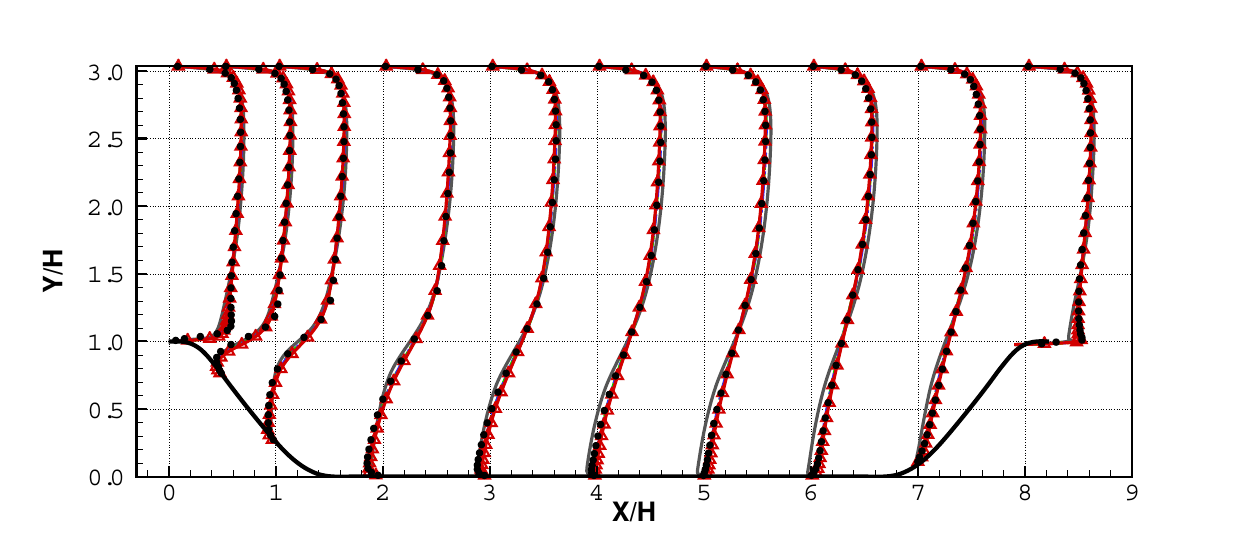}    }
        \\
        \subfigure[$6000 k/U_b^2 + x/H$]
        {\includegraphics[width=1\textwidth]{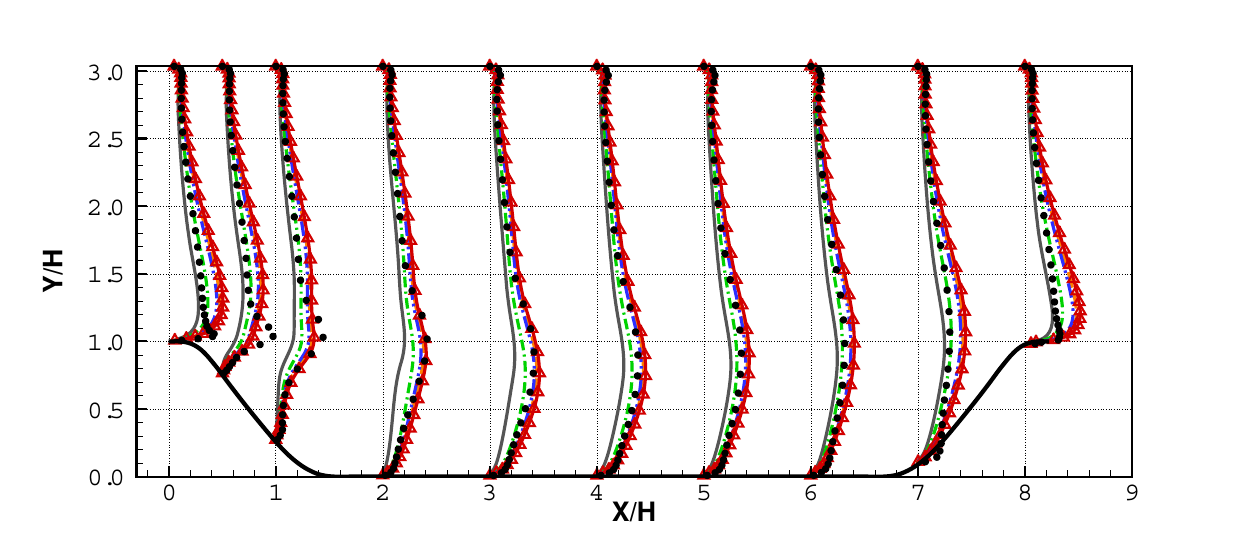}    }
        \\
        \subfigure[$15000 \tau_{12}/U_b^2 + x/H$]
        {\includegraphics[width=1\textwidth]{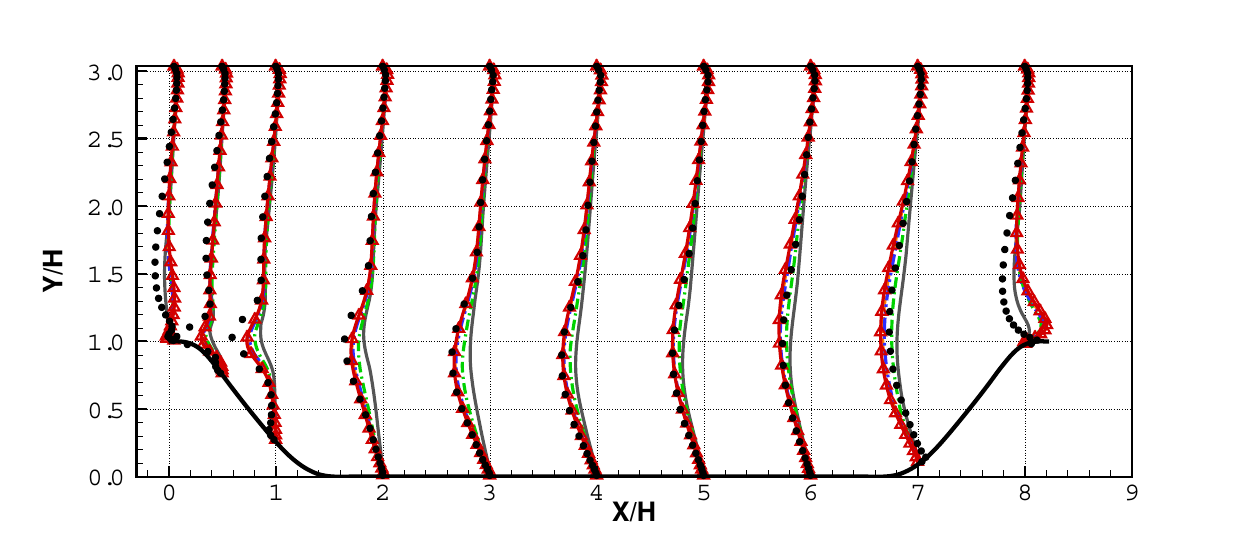}    }
        \caption{Estimation for various SR models for flow over Periodic hills $Re = 5,600, \alpha = 0.8$.   }
        \label{fig:PH_08}
    \end{minipage}
    \hfill
    \begin{minipage}{0.48\textwidth}
        \centering
        \subfigure[$20 U_1/U_b + x/H$]
        {\includegraphics[width=1\textwidth]{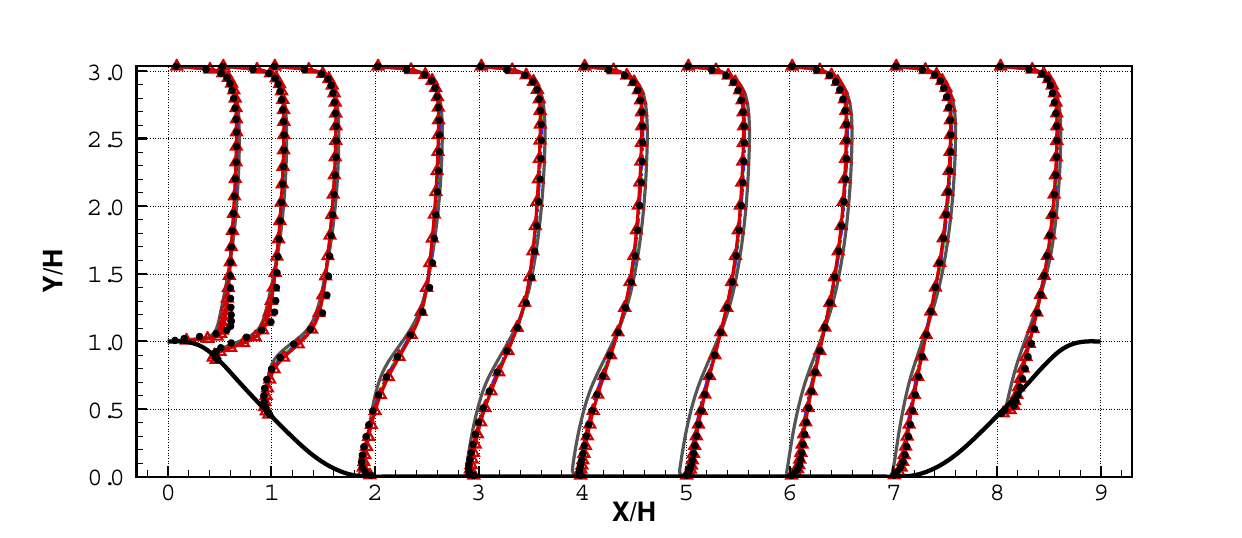}    }
        \\
        \subfigure[$6000 k/U_b^2 + x/H$]
        {\includegraphics[width=1\textwidth]{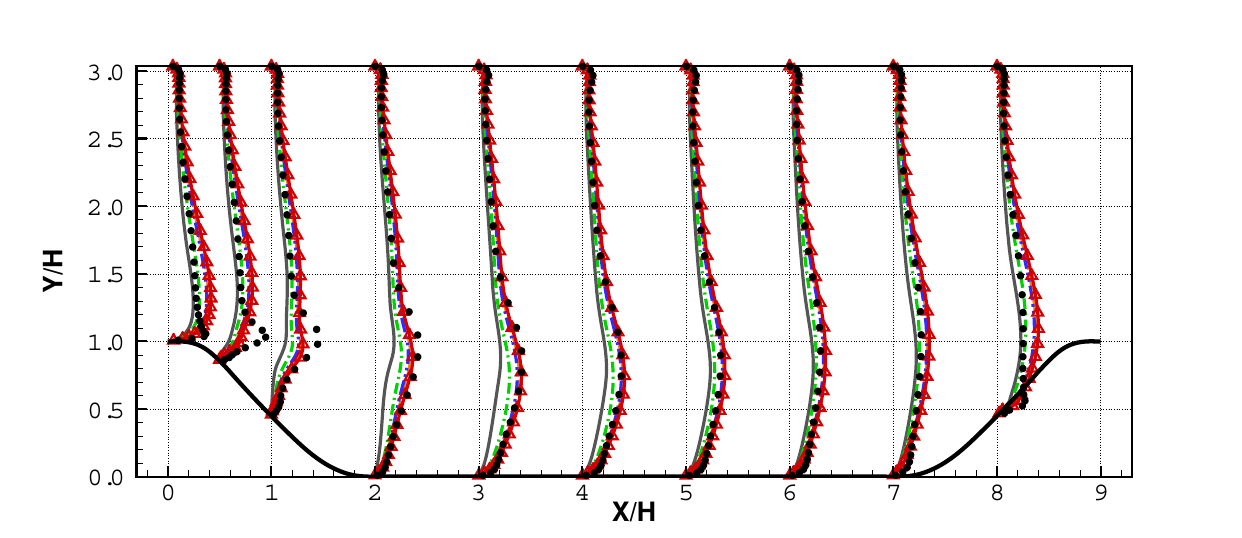}    }
        \\
        \subfigure[$15000 \tau_{12}/U_b^2 + x/H$]
        {\includegraphics[width=1\textwidth]{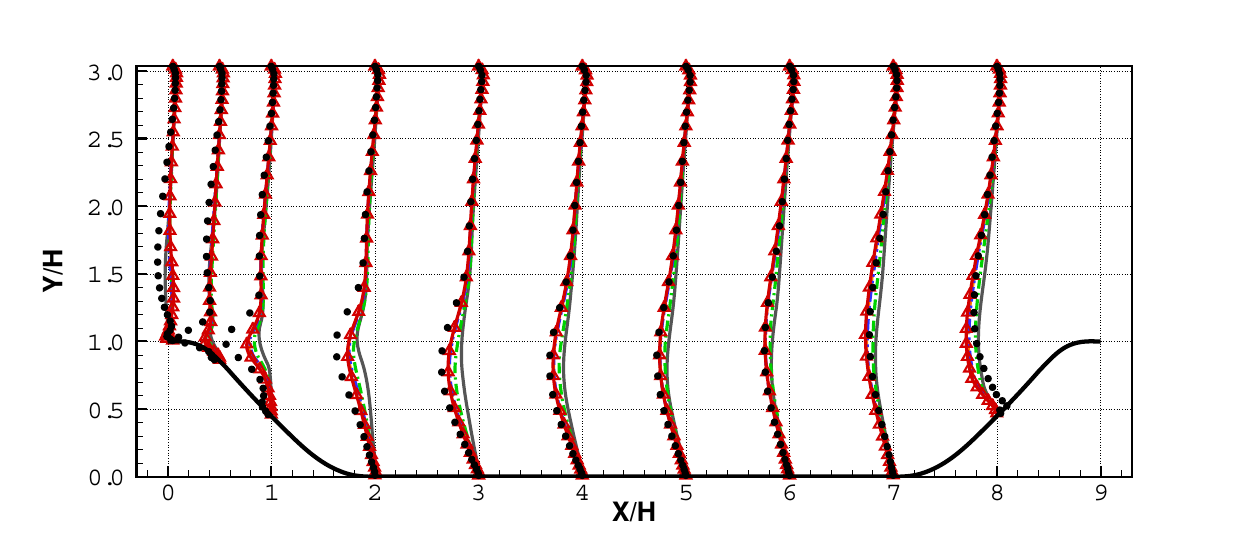}    }
        \caption{Estimation for various SR models for flow over Periodic hills $Re = 5,600, \alpha = 1.0$.    }
        \label{fig:PH_10}    
    \end{minipage}
\end{figure*}
\begin{figure*} [htbp]
    \centering
    \begin{minipage}{0.48\textwidth}
        \centering
        \subfigure[$20 U_1/U_b + x/H$]
        {\includegraphics[width=1\textwidth]{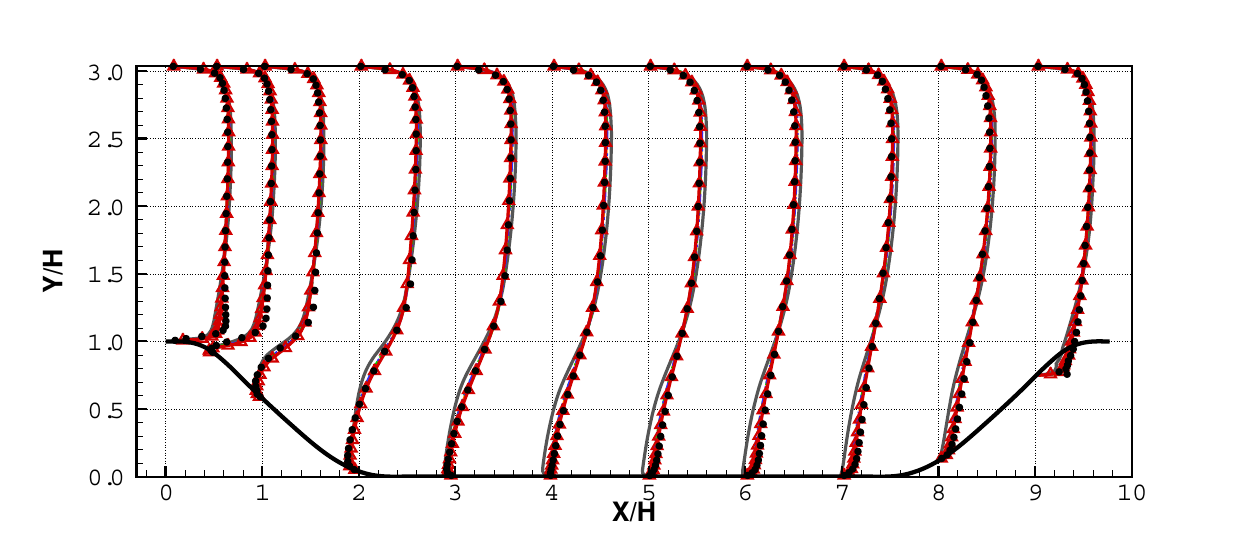}    }
        \\
        \subfigure[$6000 k/U_b^2 + x/H$]
        {\includegraphics[width=1\textwidth]{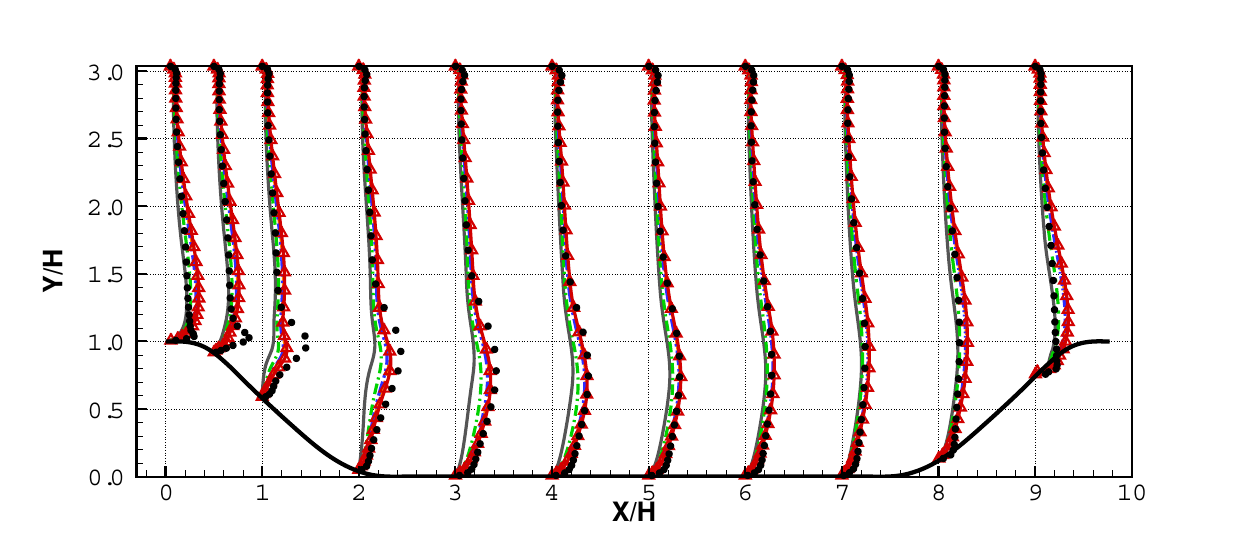}    }
        \\
        \subfigure[$15000 \tau_{12}/U_b^2 + x/H$]
        {\includegraphics[width=1\textwidth]{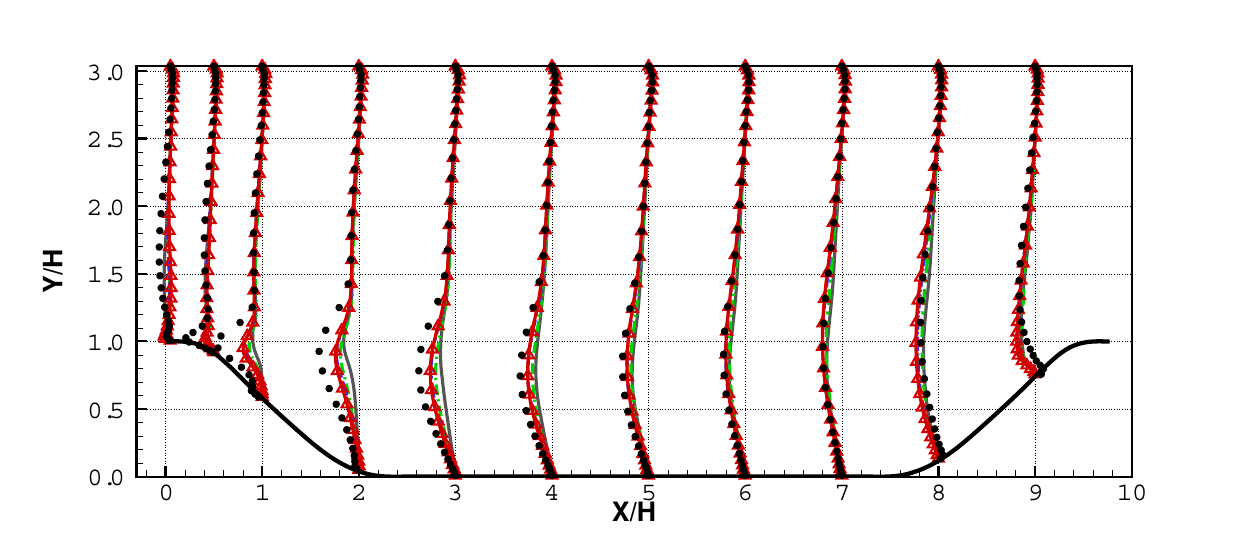}    }
        \caption{Estimation for various SR models for flow over Periodic hills $Re = 5,600, \alpha = 1.2$.   }
        \label{fig:PH_12}
    \end{minipage}
    \hfill
    \begin{minipage}{0.48\textwidth}
        \centering
        \subfigure[$20 U_1/U_b + x/H$]
        {\includegraphics[width=1\textwidth]{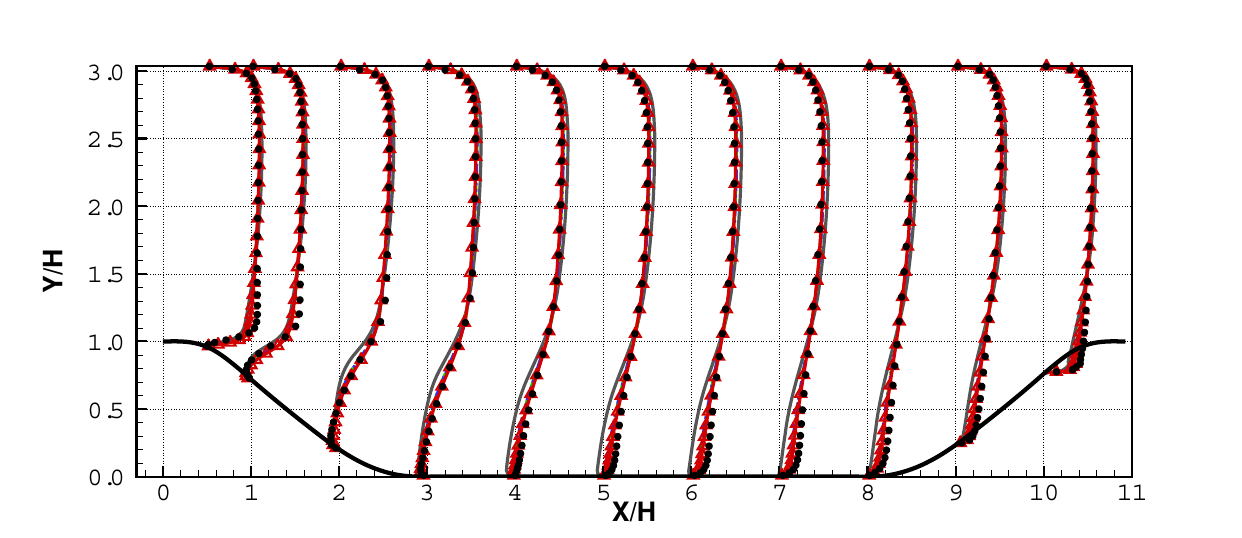}    }
        \\
        \subfigure[$6000 k/U_b^2 + x/H$]
        {\includegraphics[width=1\textwidth]{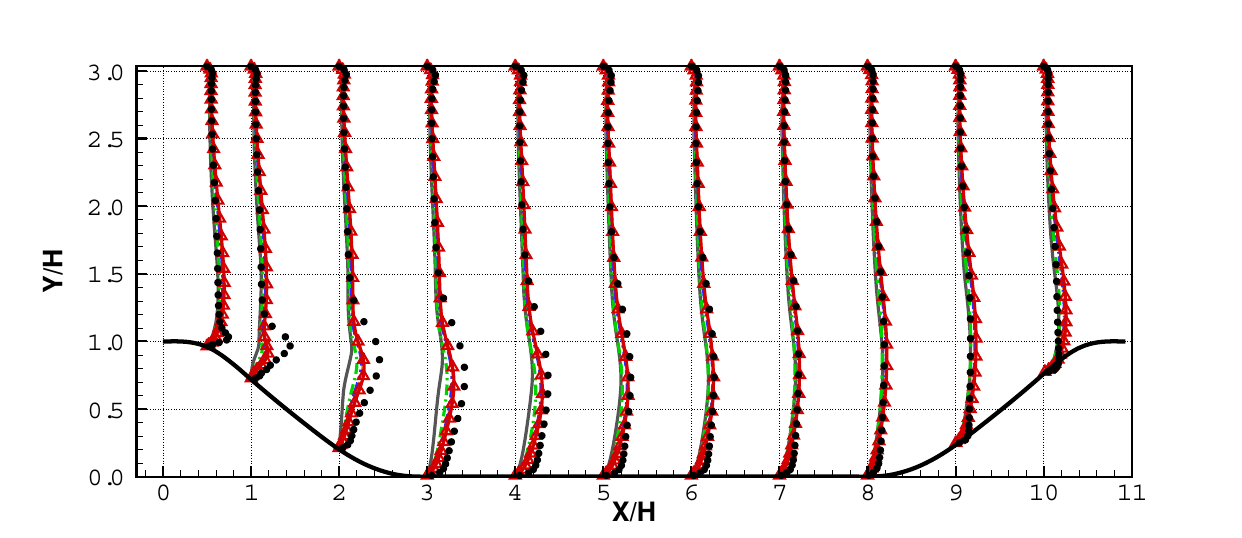}    }
        \\
        \subfigure[$15000 \tau_{12}/U_b^2 + x/H$]
        {\includegraphics[width=1\textwidth]{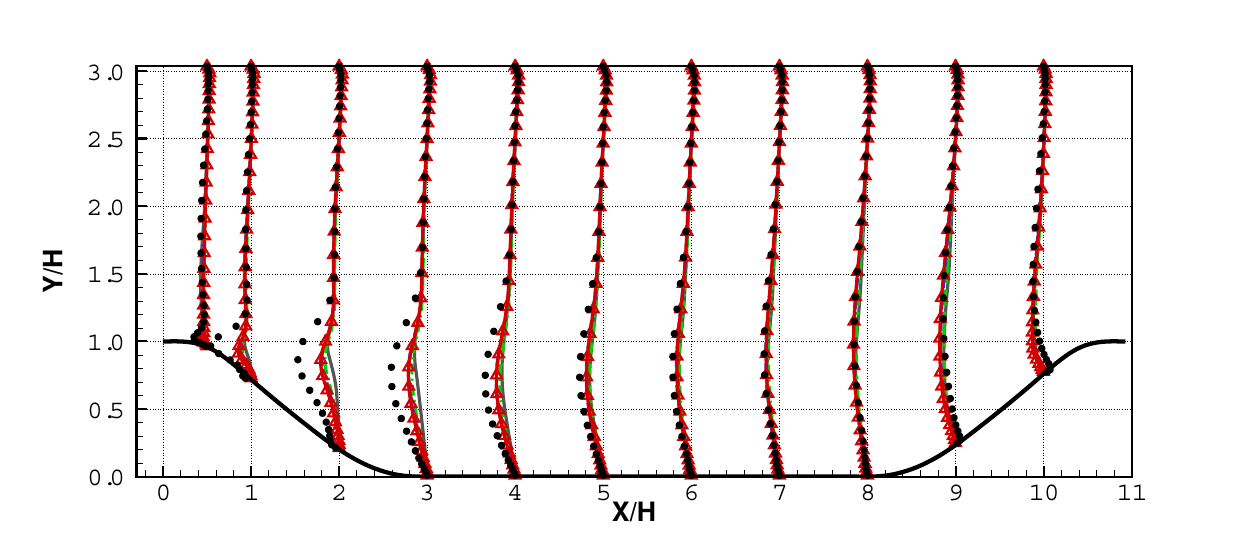}    }
        \caption{Estimation for various SR models for flow over Periodic hills $Re = 5,600, \alpha = 1.5$.    }
        \label{fig:PH_15}    
    \end{minipage}
\end{figure*}
\begin{figure*} [htbp]
    \centering
    \begin{minipage}{0.48\textwidth}
        \centering         
        \subfigure[$U_1/U_b + x/H$]
        {\includegraphics[width=1\textwidth]{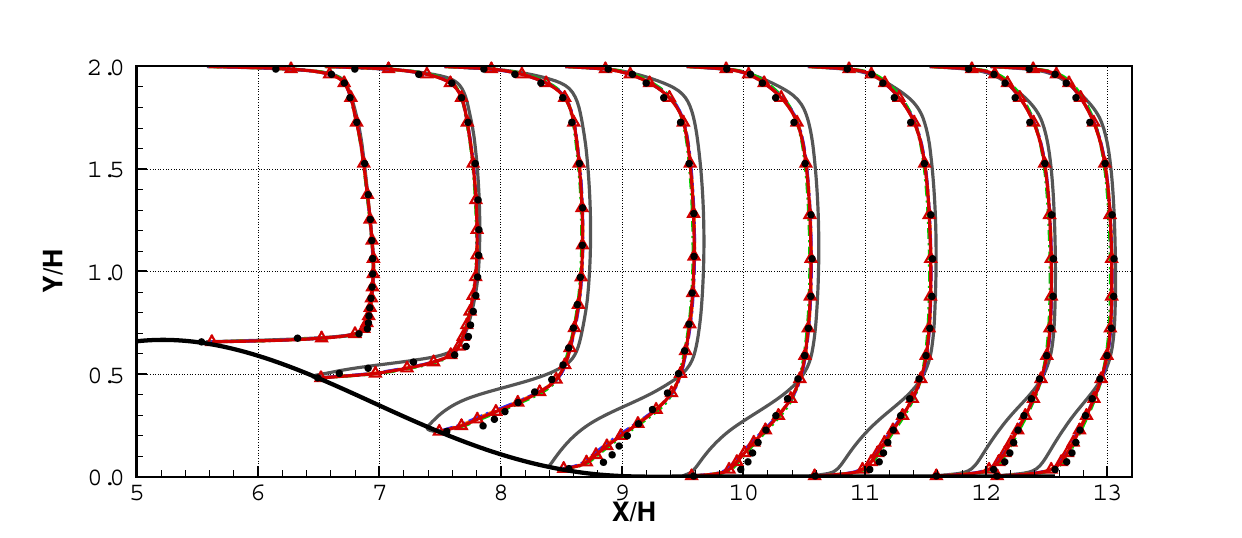}    }
        \\
        \subfigure[$12k/U_b^2 + x/H$]
        {\includegraphics[width=1\textwidth]{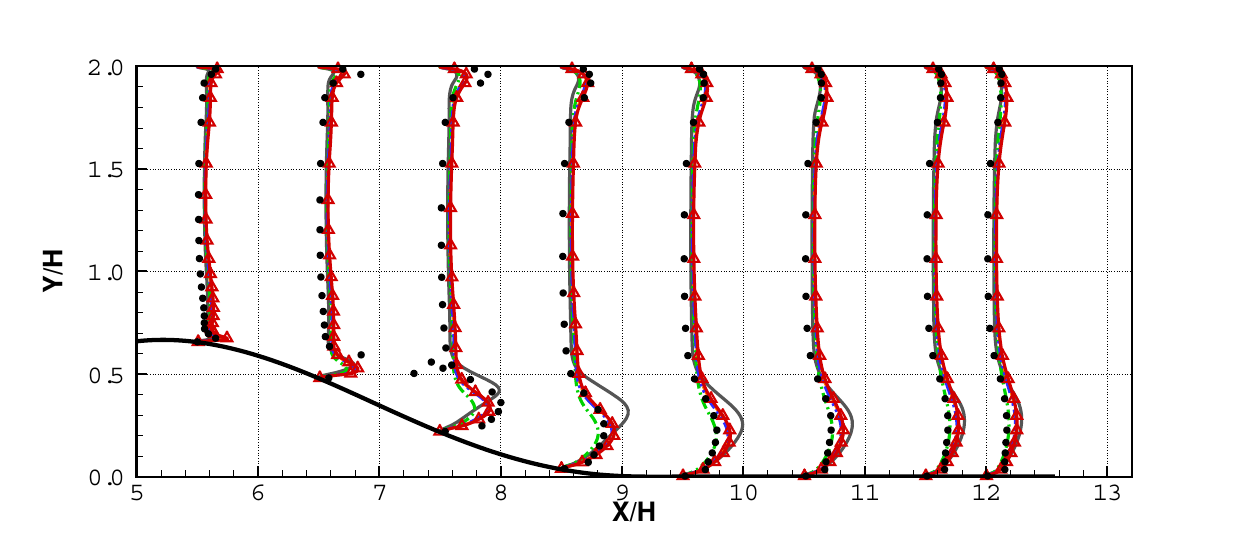}    }
        \\
        \subfigure[$25\tau_{12}/U_b^2 + x/H$]
        {\includegraphics[width=1\textwidth]{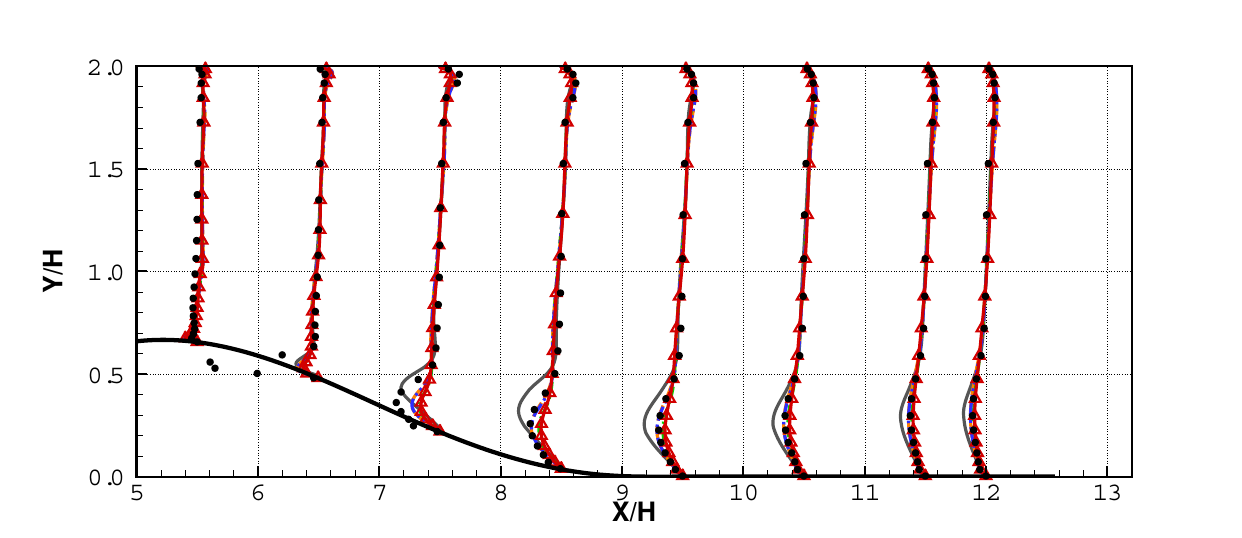}    }
        \caption{Estimation for various SR models for flow over converging-diverging channel $Re = 12,600$.  }
        \label{fig:CD}   
    \end{minipage}
    \hfill
    \begin{minipage}{0.48\textwidth}
        \centering
        \subfigure[$U_1/U_b + x/H$]
        {\includegraphics[width=1\textwidth]{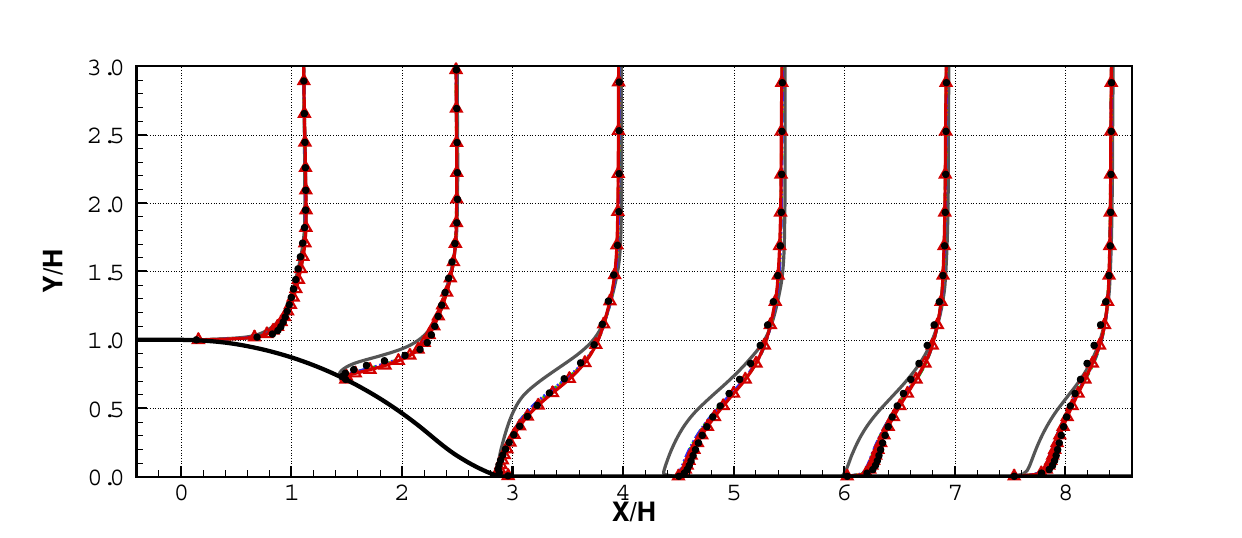}    }
        \\
        \subfigure[$12k/U_b^2 + x/H$]
        {\includegraphics[width=1\textwidth]{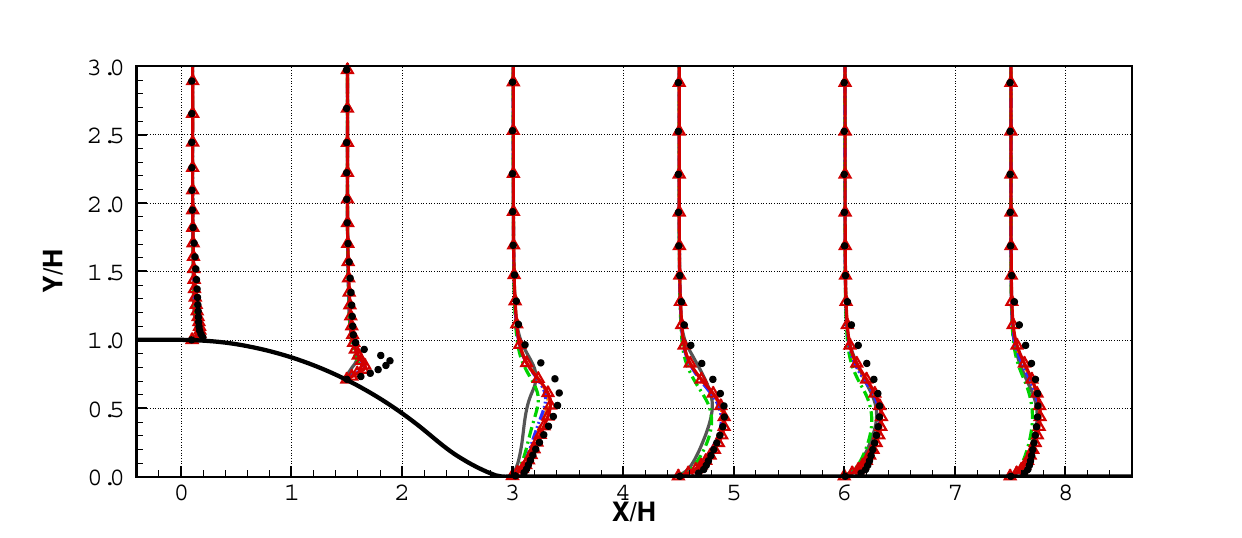}    }
        \\
        \subfigure[$20\tau_{12}/U_b^2 + x/H$]
        {\includegraphics[width=1\textwidth]{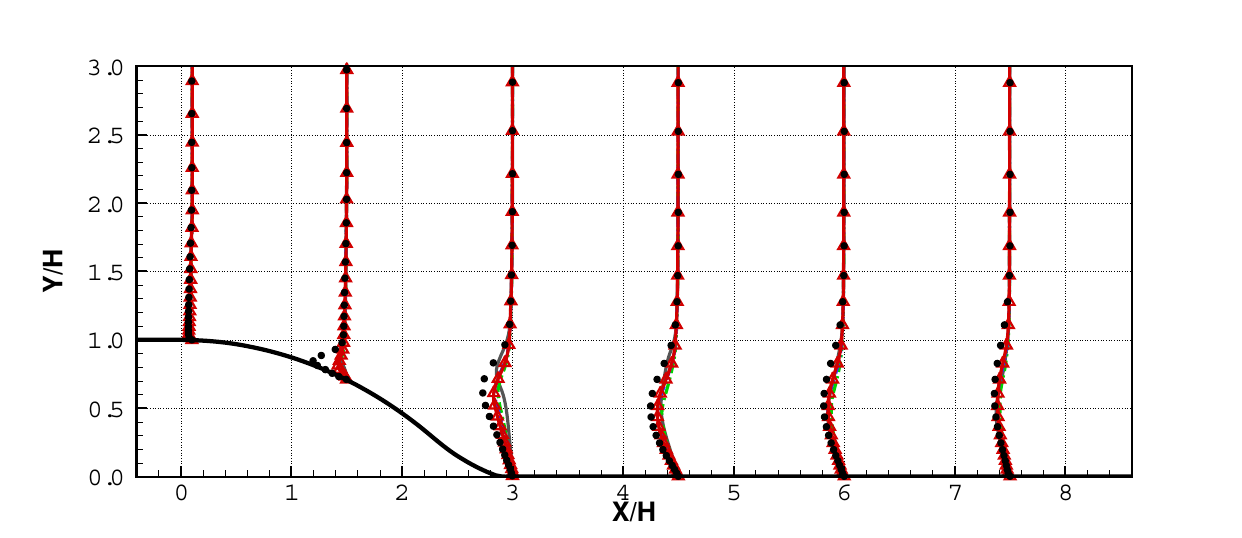}    }
        \caption{Estimation for various SR models for flow over curved back-facing step $Re = 13,700$.   }
        \label{fig:CBF}        
    \end{minipage}
\end{figure*}

\section{Prompts for AutoTurb}
\label{app3}

\begin{tcolorbox}[colback=white, colframe=black, boxrule=0.5mm, arc=3mm, title=Initialization Prompt (IP)]
    
    \sethlcolor{brighterlightblue}
    \hl{Given two input variables l1 and l2, please help me design two non-linear expressions to approximate real-world equations.} \\
    \sethlcolor{lightred}
    \hl{The format must be as follows: <start> bDelta\_ = T1*(...) + T2*(...) + T3*(...);  kDeficit\_ =  (T1*(...) + T2*(...) + T3*(...)) \&\& tgradU()*2.0*k\_; <end> In this expression, replace '...' with appropriate functions of variables l1 and l2 and keep T1*, T2* and T3*.  Do not modify \&\& tgradU()*2.0*k\_. Ensure the formulation is compatible with C++ programming standards.} \\
\end{tcolorbox}

\begin{tcolorbox}[colback=white, colframe=black, boxrule=0.5mm, arc=3mm, title=Evolution Prompt (EP1)] 
    \sethlcolor{brighterlightblue}
    \hl{Given two input variables l1 and l2, please help me design two non-linear expressions to approximate real-world equations.} \\
    \sethlcolor{lightgrey}
    \hl{I have 2 pairs of expressions as follows: \\
    <Equation 1> \\
    <Equation 2> } \\
    \sethlcolor{lightgreen}
    \hl{Please help me create a new expression that has a totally different form from the given ones.}  \\
    \sethlcolor{lightred}
    \hl{The format must be as follows: <start> bDelta\_ = T1*(...) + T2*(...) + T3*(...);  kDeficit\_ =  (T1*(...) + T2*(...) + T3*(...)) \&\& tgradU()*2.0*k\_; <end> In this expression, replace '...' with appropriate functions of variables l1 and l2 and keep T1*, T2* and T3*.  Do not modify \&\& tgradU()*2.0*k\_. Ensure the formulation is compatible with C++ programming standards.} \\
\end{tcolorbox}

\begin{tcolorbox}[colback=white, colframe=black, boxrule=0.5mm, arc=3mm, title=Evolution Prompt (EP2)] 
    \sethlcolor{brighterlightblue}
    \hl{Given two input variables l1 and l2, please help me design two non-linear expressions to approximate real-world equations.} \\

    \sethlcolor{lightgrey}
    \hl{I have 2 pairs of expressions as follows: \\
    <Equation 1> \\
    <Equation 2> } \\
    \sethlcolor{lightgreen}
    \hl{Please help me create a new expression that is motivated by the given ones.}  \\
    \sethlcolor{lightred}
    \hl{The format must be as follows: <start> bDelta\_ = T1*(...) + T2*(...) + T3*(...);  kDeficit\_ =  (T1*(...) + T2*(...) + T3*(...)) \&\& tgradU()*2.0*k\_; <end> In this expression, replace '...' with appropriate functions of variables l1 and l2 and keep T1*, T2* and T3*.  Do not modify \&\& tgradU()*2.0*k\_. Ensure the formulation is compatible with C++ programming standards.} \\
\end{tcolorbox}

\begin{tcolorbox}[colback=white, colframe=black, boxrule=0.5mm, arc=3mm, title=Evolution Prompt (EP3)] 
    \sethlcolor{brighterlightblue}
    \hl{Given two input variables l1 and l2, please help me design two non-linear expressions to approximate real-world equations.} \\
    \sethlcolor{lightgrey}
    \hl{I have one expression as follows: \\
    <Equation 1> } \\
    \sethlcolor{lightgreen}
    \hl{Please help me create a new equation that is a revision of the given one.}  \\
    \sethlcolor{lightred}
    \hl{The format must be as follows: <start> bDelta\_ = T1*(...) + T2*(...) + T3*(...);  kDeficit\_ =  (T1*(...) + T2*(...) + T3*(...)) \&\& tgradU()*2.0*k\_; <end> In this expression, replace '...' with appropriate functions of variables l1 and l2 and keep T1*, T2* and T3*.  Do not modify \&\& tgradU()*2.0*k\_. Ensure the formulation is compatible with C++ programming standards.} \\
\end{tcolorbox}

\begin{tcolorbox}[colback=white, colframe=black, boxrule=0.5mm, arc=3mm, title=Evolution Prompt (EP4)] 
    \sethlcolor{brighterlightblue}
    \hl{Given two input variables l1 and l2, please help me design two non-linear expressions to approximate real-world equations.} \\
    \sethlcolor{lightgrey}
    \hl{I have one expression as follows: \\
    <Equation 1> } \\
    \sethlcolor{lightgreen}
    \hl{Please help me create a new equation that has different parameter settings of the given one.}  \\
    \sethlcolor{lightred}
    \hl{The format must be as follows: <start> bDelta\_ = T1*(...) + T2*(...) + T3*(...);  kDeficit\_ =  (T1*(...) + T2*(...) + T3*(...)) \&\& tgradU()*2.0*k\_; <end> In this expression, replace '...' with appropriate functions of variables l1 and l2 and keep T1*, T2* and T3*.  Do not modify \&\& tgradU()*2.0*k\_. Ensure the formulation is compatible with C++ programming standards.} \\
\end{tcolorbox}

\bibliographystyle{elsarticle-num} 
\bibliography{reference}




% %% For numbered reference style

% \end{thebibliography}
\end{document}